\documentclass[pdflatex,sn-nature,oneside]{sn-jnl}
\usepackage{graphicx}
\usepackage{multirow}
\usepackage{amsmath,amssymb,amsfonts}
\usepackage{amsthm}
\usepackage{mathrsfs}
\usepackage[title]{appendix}
\usepackage{xcolor}
\usepackage{textcomp}
\usepackage{manyfoot}
\usepackage{booktabs}
\usepackage{algorithm}
\usepackage{algorithmicx}
\usepackage{algpseudocode}
\usepackage{listings}
\usepackage{makecell}
\usepackage{float}
\usepackage{array}
\newcolumntype{C}[1]{>{\centering\arraybackslash}p{#1}}

\begin{document}

\title[Topological-Insulator Gate Stacks]{\begin{minipage}{\textwidth}\centering
     Topological-Insulator Heterophase Gate Stacks \\ for Transistor Electrostatics
\end{minipage}}

\author[1]{\fnm{Minuk} \sur{Song}}
\equalcont{These authors contributed equally to this work.}

\author[2]{\fnm{Jiwan} \sur{Kim}}
\equalcont{These authors contributed equally to this work.}

\author[3,4]{\fnm{Md Gius} \sur{Uddin}}
\equalcont{These authors contributed equally to this work.}

\author[1]{\fnm{Wonseok} \sur{Kim}}
\equalcont{These authors contributed equally to this work.}

\author[1]{\fnm{Jihun} \sur{Park}}

\author[5]{\fnm{Han Uk} \sur{Lee}}

\author[5]{\fnm{Dong Won} \sur{Jeon}}

\author[5,6]{\fnm{Dohyung} \sur{Lee}}

\author[7]{\fnm{Lide} \sur{Yao}}

\author[8]{\fnm{Jouko} \sur{Lahtinen}}

\author[1]{\fnm{Gyunghyun} \sur{Jang}}

\author[1]{\fnm{Soohyun} \sur{Min}}

\author[1]{\fnm{Yonas Tsegaye} \sur{Megra}}

\author[3]{\fnm{Xiaoqi} \sur{Cui}}

\author[9]{\fnm{Seungwoo} \sur{Choi}}

\author[10]{\fnm{Yunyun} \sur{Dai}}

\author[11]{\fnm{Sang Hoon} \sur{Chae}}

\author[9]{\fnm{Keun Su} \sur{Kim}}

\author[1]{\fnm{Dong-Ho} \sur{Kang}}

\author[1]{\fnm{Hyeon-Jin} \sur{Shin}}

\author[6,12]{\fnm{Wooseok} \sur{Song}}

\author[13]{\fnm{Seth Ariel} \sur{Tongay}}

\author[14]{\fnm{Chul-Ho} \sur{Lee}}

\author[15]{\fnm{Manish} \sur{Chhowalla}}

\author*[5]{\fnm{Sung Beom} \sur{Cho}}\email{sungcho@skku.edu}

\author*[3]{\fnm{Zhipei} \sur{Sun}}\email{zhipei.sun@aalto.fi}

\author*[2]{\fnm{Kibog} \sur{Park}}\email{kibogpark@unist.ac.kr}

\author*[1]{\fnm{Hoon Hahn} \sur{Yoon}}\email{hoonhahnyoon@gist.ac.kr}

\affil[1]{\orgdiv{Department of Semiconductor Engineering}, \orgname{Gwangju Institute of Science and Technology}, \orgaddress{\city{Gwangju}, \postcode{61005}, \country{Republic of Korea}}}

\affil[2]{\orgdiv{Department of Physics}, \orgname{Ulsan National Institute of Science and Technology}, \orgaddress{\city{Ulsan}, \postcode{44919}, \country{Republic of Korea}}}

\affil[3]{\orgdiv{Department of Electronics and Nanoengineering}, \orgname{Aalto University}, \orgaddress{\city{Espoo}, \postcode{02150}, \country{Finland}}}

\affil[4]{\orgname{VTT Technical Research Centre of Finland Ltd.}, \orgaddress{\city{Espoo}, \postcode{02150}, \country{Finland}}}

\affil[5]{\orgdiv{School of Advanced Materials Science and Engineering}, \orgname{Sungkyunkwan University}, \orgaddress{\city{Suwon}, \postcode{16419}, \country{Republic of Korea}}}

\affil[6]{\orgdiv{Thin Film Materials Research Center}, \orgname{Korea Research Institute of Chemical Technology}, \orgaddress{\city{Daejeon}, \postcode{34114}, \country{Republic of Korea}}}

\affil[7]{\orgdiv{OtaNano-Nanomicroscopy Center}, \orgname{Aalto University}, \orgaddress{\city{Espoo}, \postcode{02150}, \country{Finland}}}

\affil[8]{\orgdiv{Department of Applied Physics}, \orgname{Aalto University}, \orgaddress{\city{Espoo}, \postcode{02150}, \country{Finland}}}

\affil[9]{\orgdiv{Department of Physics}, \orgname{Yonsei University}, \orgaddress{\city{Seoul}, \postcode{03722}, \country{Republic of Korea}}}

\affil[10]{\orgdiv{School of Interdisciplinary Science}, \orgname{Beijing Institute of Technology}, \orgaddress{\city{Beijing}, \postcode{100081}, \country{China}}}

\affil[11]{\orgdiv{School of Electrical and Electronic Engineering}, \orgname{Nanyang Technological University}, \orgaddress{\city{Singapore}, \postcode{639798}, \country{Singapore}}}

\affil[12]{\orgdiv{School of Electronic and Electrical Engineering}, \orgname{Sungkyunkwan University}, \orgaddress{\city{Suwon}, \postcode{16419}, \country{Republic of Korea}}}

\affil[13]{\orgdiv{Department of Materials Science and Engineering, School for Engineering of Matter, Transport and Energy}, \orgname{Arizona State University}, \orgaddress{\city{Tempe}, \postcode{AZ 85287}, \country{United States}}}

\affil[14]{\orgdiv{Department of Electrical and Computer Engineering}, \orgname{Seoul National University}, \orgaddress{\city{Seoul}, \postcode{08826}, \country{Republic of Korea}}}

\affil[15]{\orgdiv{Department of Materials Science \& Metallurgy}, \orgname{University of Cambridge}, \orgaddress{\city{Cambridge}, \postcode{CB3 0FA}, \country{United Kingdom}}}

\abstract{
Conventional gate-stack scaling reduces dielectric thickness and increases permittivity while largely treating the position and electronic character of the gate-side screening boundary as fixed. Here we show that this boundary can be engineered by converting the surface of the topological insulator Bi\textsubscript{2}Se\textsubscript{3} into insulating high-$\kappa$ BiF\textsubscript{3}. Position-resolved calculations reveal a gap-opened immediate amorphous-BiF\textsubscript{3}/crystalline-Bi\textsubscript{2}Se\textsubscript{3} interface and a reconstructed gap-closed Bi\textsubscript{2}Se\textsubscript{3}-derived state in the adjacent subinterface layer, accompanied by a localized interfacial dipole. Capacitor measurements independently resolve a finite series response consistent with electronic compressibility at this boundary; this response lowers the nominal stack capacitance. Despite this capacitance penalty, MoS\textsubscript{2} transistors with closely matched BiF\textsubscript{3} thicknesses and a common BiF\textsubscript{3}/MoS\textsubscript{2} channel-side material interface exhibit near-thermionic-limit switching, negligible hysteresis and substantially weaker subthreshold drain-bias dependence than BiF\textsubscript{3}-only controls. These results identify the position and electronic character of the gate-side screening boundary as additional design variables for transistor electrostatics beyond nominal dielectric capacitance.
}

\maketitle

\section{Introduction}
\label{sec1}

Gate stacks for low-voltage transistors must provide effective channel control while maintaining dielectric isolation and a stable electrical response. Gate-stack scaling is conventionally described by dielectric thickness and permittivity, yet the applied potential is partitioned across the bulk insulator and finite screening and interfacial responses at conductor/dielectric and dielectric/semiconductor boundaries. As the equivalent oxide thickness ($EOT$) approaches the subnanometre regime, contributions arising from electrode screening, interfacial bonding, polarization, and chemical or structural reconstruction need not scale with the physical dielectric thickness and can therefore constitute an increasing fraction of the total electrostatic response~\cite{pourfath2026device,cao2023future,lau2023dielectrics,kim2025gate,jung2025advances}. Relating gate-stack capacitance to subthreshold switching and drain-bias sensitivity therefore requires consideration of the electronic boundaries at which screening charge is established, in addition to the properties of the bulk dielectric.

Two-dimensional (2D) van der Waals (vdW) semiconductors provide a stringent setting in which to expose this distinction. Their atomically thin bodies suppress semiconductor-body electrostatic length scales, whereas their chemically inert surfaces make conventional three-dimensional high-$\kappa$ integration difficult and susceptible to interfacial disorder, fixed charge, trap states, and carrier scattering~\cite{shin20252d,yoon2025enabling,jung2026advances}. Hexagonal boron nitride encapsulation, seeded atomic-layer deposition, native dielectrics, and emerging vdW high-$\kappa$ materials have therefore focused primarily on improving the dielectric and its interface with the semiconductor~\cite{osada2012two,lee2015highly,li2019uniform,li2020native,xu2023scalable,su2026high}. These advances leave a complementary question largely open: whether the gate-side screening boundary itself can be deliberately created, displaced, and electronically reconstructed independently of the semiconductor-side interface.

Topological insulators provide a route to this form of boundary engineering. Their bulk electronic structure supports boundary-localized states whose spatial distribution can be displaced or reconstructed when the surface environment is modified~\cite{chen2009experimental,moore2010birth,pesin2012spintronics,yue2024topological}. Such states have predominantly been considered as transport channels~\cite{sun2021topological,breunig2022opportunities}. By contrast, their use as an electrostatically active boundary of a gate stack remains largely unexplored. Because the surface of Bi\textsubscript{2}Se\textsubscript{3} can be selectively converted into wide-gap BiF\textsubscript{3}~\cite{barton2019impact}, surface conversion provides a means to separate the chemical surface from the electronically active Bi\textsubscript{2}Se\textsubscript{3}-derived boundary beneath it. This possibility motivates a gate architecture in which the location of screening is an engineered property of the heterophase material itself rather than a fixed consequence of depositing a metal electrode onto a dielectric.

\begin{figure*}[!b]
\centering
\includegraphics[width=\textwidth]{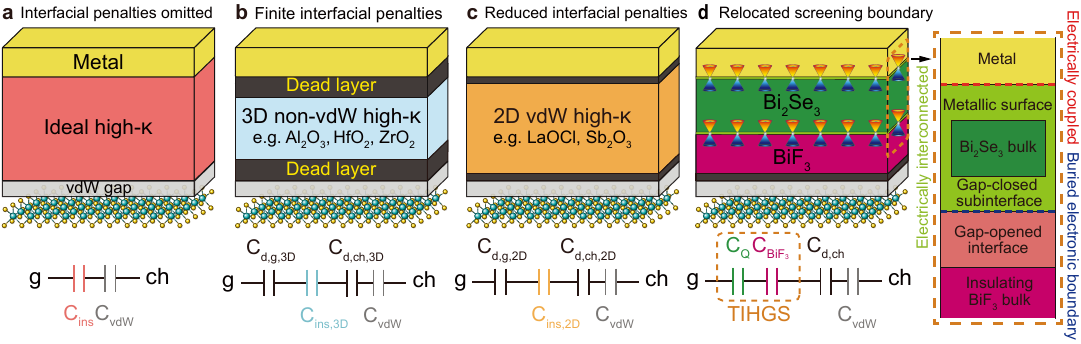}
\caption{\textbf{Screening-boundary relocation in a topological-insulator heterophase gate stack (TIHGS).}
\textbf{a--d}, Schematics and reduced equivalent-capacitance models for gate stacks with a two-dimensional (2D) van der Waals (vdW) channel. The terminals g and ch denote the gate and channel; the bias-dependent semiconductor-body capacitance is omitted. $C_{\mathrm{ins}}$, $C_{\mathrm{vdW}}$, $C_{\mathrm{d,g}}$ and $C_{\mathrm{d,ch}}$ denote the insulator, vdW-gap, effective gate-side interfacial and effective channel-side interfacial capacitances, respectively.
\textbf{a}, Idealized stack retaining $C_{\mathrm{ins}}$ and $C_{\mathrm{vdW}}$ while omitting the gate-side and channel-side interfacial-capacitance penalties.
\textbf{b,c}, Three-dimensional (3D) non-vdW \textbf{(b)} and 2D vdW \textbf{(c)} high-$\kappa$ dielectrics retaining finite interfacial contributions in series with the insulator and vdW-gap capacitances. The 3D and 2D subscripts identify the dielectric class. Interfacial layers and relative penalties are schematic.
\textbf{d}, Surface conversion of Bi\textsubscript{2}Se\textsubscript{3} forms insulating BiF\textsubscript{3} facing the channel, while residual conducting Bi\textsubscript{2}Se\textsubscript{3} remains electrically coupled to the gate electrode. The right-hand schematic distinguishes the gap-opened amorphous-BiF\textsubscript{3}/crystalline-Bi\textsubscript{2}Se\textsubscript{3} chemical interface from the adjacent gap-closed Bi\textsubscript{2}Se\textsubscript{3}-derived subinterface electronic boundary. $C_Q$ represents the finite effective electronic-compressibility response of this buried boundary, in series with the BiF\textsubscript{3} dielectric capacitance ($C_{\mathrm{BiF_3}}$), $C_{\mathrm{d,ch}}$ and $C_{\mathrm{vdW}}$. Cone symbols schematically represent Bi\textsubscript{2}Se\textsubscript{3}-derived Dirac-like boundary states. The dashed box in the equivalent circuit identifies the TIHGS element comprising $C_Q$ and $C_{\mathrm{BiF_3}}$ in series.}
\label{Fig:TIHGS_Concept}
\end{figure*}

Here we investigate this gate-side boundary condition using a topological-insulator heterophase gate stack (TIHGS) formed by selective conversion of Bi\textsubscript{2}Se\textsubscript{3}. Fluorination creates an insulating amorphous BiF\textsubscript{3} phase while retaining electrically conducting crystalline Bi\textsubscript{2}Se\textsubscript{3} underneath. Position-resolved electronic-structure calculations show that the immediate heterophase interface becomes gap-opened whereas an electronically active, gap-closed Bi\textsubscript{2}Se\textsubscript{3}-derived state is reconstructed in the adjacent subinterface layer; a localized interface dipole accompanies this reconstruction. Independent capacitor measurements resolve a finite series response consistent with electronic compressibility at the buried boundary, lowering the nominal stack capacitance. Nevertheless, TIHGS-gated MoS\textsubscript{2} transistors exhibit near-thermionic-limit switching, negligible hysteresis and reduced drain-bias sensitivity relative to a BiF\textsubscript{3}-only control with closely matched dielectric thickness and a common channel-side material interface. Together, these measurements connect conversion-induced electronic reconstruction, gate-stack response and transistor characteristics, identifying the gate-side screening boundary as a design parameter complementary to dielectric thickness and permittivity.

\section{Results and Discussion}
\label{sec2}

\begin{figure*}[!b]
\centering
\includegraphics[width=\textwidth]{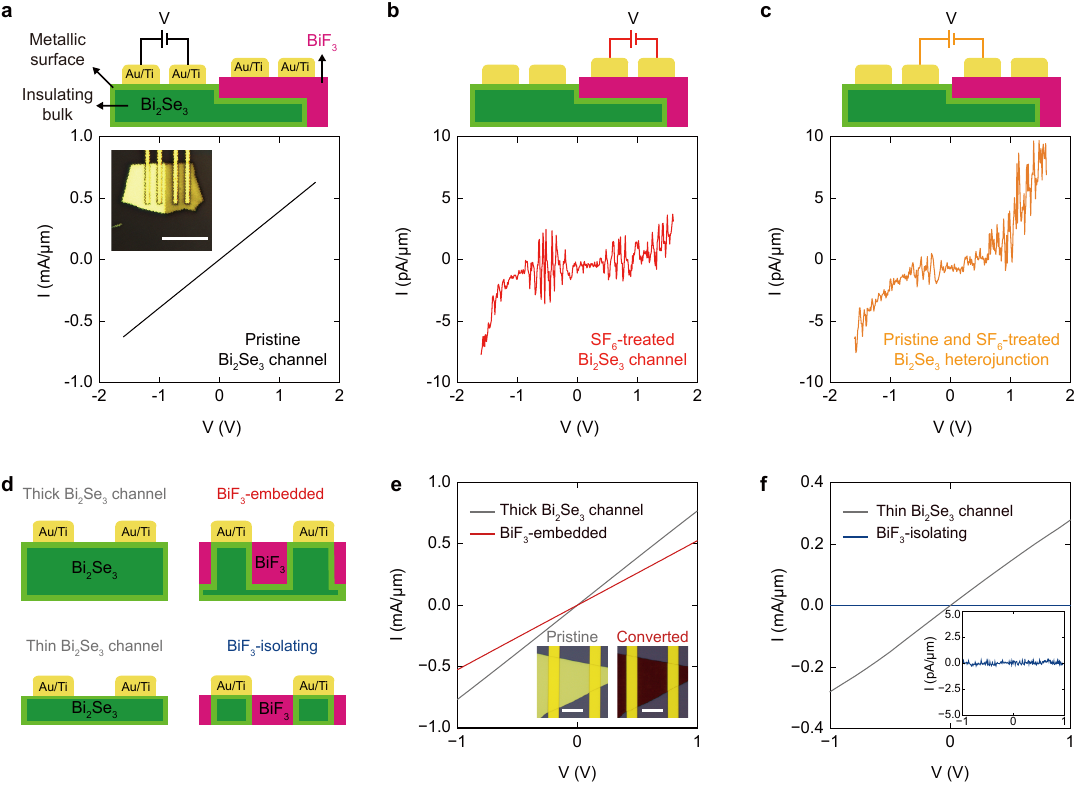}
\caption{\textbf{Electrical isolation and buried conduction in the topological-insulator heterophase gate stack.}
\textbf{a--c}, Device schematics (top) and corresponding current--voltage ($I$--$V$) characteristics (bottom) of \textbf{a}, pristine Bi\textsubscript{2}Se\textsubscript{3}; \textbf{b}, SF\textsubscript{6}-converted Bi\textsubscript{2}Se\textsubscript{3}; and \textbf{c}, a lateral pristine/converted heterojunction. The converted region suppresses conduction from the mA~$\mu$m$^{-1}$ to the pA~$\mu$m$^{-1}$ range. Scale bar in \textbf{a}, 10~$\mu$m.
\textbf{d}, Schematics illustrating partial surface conversion of a Bi\textsubscript{2}Se\textsubscript{3} flake thicker than the conversion depth and complete conversion of a thinner flake.
\textbf{e}, $I$--$V$ characteristics of a thick Bi\textsubscript{2}Se\textsubscript{3} channel before and after partial surface conversion, showing persistent conduction through the residual Bi\textsubscript{2}Se\textsubscript{3}. Insets, corresponding optical micrographs. Scale bars, 20~$\mu$m.
\textbf{f}, $I$--$V$ characteristics of a thin Bi\textsubscript{2}Se\textsubscript{3} channel before and after complete conversion, showing electrical isolation after fluorination. Inset, current on the pA scale after conversion.}
\label{Fig:Electrical_Realization}
\end{figure*}

Figure~\ref{Fig:TIHGS_Concept} contrasts conventional gate-stack electrostatics with screening-boundary relocation in the TIHGS. In a conventional high-$\kappa$ stack, the gate-side interfacial capacitance ($C_{\mathrm{d,g}}$), dielectric capacitance, channel-side interfacial capacitance ($C_{\mathrm{d,ch}}$), and, for a 2D channel, the van der Waals-gap capacitance ($C_{\mathrm{vdW}}$) contribute in series. Of these terms, $C_{\mathrm{d,g}}$ is central to the present work: it represents the finite electrostatic response of the gate-electrode/dielectric boundary arising from electrode screening and interface-specific electronic, bonding, polarization, and chemical or structural effects. Unlike the bulk dielectric capacitance, which increases as the physical dielectric thickness is reduced, this gate-side boundary contribution need not scale proportionally and can therefore account for an increasing fraction of the total electrostatic response at small $EOT$~\cite{lau2023dielectrics,cao2023future,jung2025advances,kim2025gate,pourfath2026device}. By comparison, $C_{\mathrm{d,ch}}$ denotes the local dielectric/semiconductor interfacial response, and $C_{\mathrm{vdW}}$ denotes the geometric contribution associated with the vdW separation at a 2D semiconductor interface. The bias-dependent semiconductor-body response is outside the gate-stack contribution defined here and is omitted from Fig.~\ref{Fig:TIHGS_Concept} for clarity. The resulting design question is therefore whether the gate-side screening boundary represented by $C_{\mathrm{d,g}}$ must remain fixed at the external metal/dielectric interface.

In the TIHGS, selective surface conversion creates insulating amorphous BiF\textsubscript{3} (a-BiF\textsubscript{3}) while residual crystalline Bi\textsubscript{2}Se\textsubscript{3} (c-Bi\textsubscript{2}Se\textsubscript{3}) remains electrically coupled to the gate electrode. This architecture displaces the gate-side electrostatic termination from a conventional externally formed metal/dielectric boundary to a heterophase-induced buried electronic boundary. Importantly, the chemical conversion front and the electronic screening boundary need not coincide atomically. The schematic in Fig.~\ref{Fig:TIHGS_Concept}d distinguishes the gap-opened a-BiF\textsubscript{3}/c-Bi\textsubscript{2}Se\textsubscript{3} interface from a reconstructed gap-closed electronic state in the adjacent c-Bi\textsubscript{2}Se\textsubscript{3} subinterface layer. We therefore use ``buried screening boundary'' to denote this electronically active heterophase-associated boundary region rather than an ideal atomically sharp conductor/dielectric plane. In the reduced series-capacitance model, its finite electronic response is represented as
\[
\frac{1}{C_{\mathrm{g,TIHGS}}}
=
\frac{1}{C_{\mathrm{TIHGS}}}
+
\frac{1}{C_{\mathrm{d,ch}}}
+
\frac{1}{C_{\mathrm{vdW}}}
=
\frac{1}{C_Q}
+
\frac{1}{C_{\mathrm{BiF_3}}}
+
\frac{1}{C_{\mathrm{d,ch}}}
+
\frac{1}{C_{\mathrm{vdW}}},
\]
where $C_Q$ denotes the effective electronic-compressibility response associated with the buried electronic boundary. Here, $C_{\mathrm{g,TIHGS}}$ includes the channel-side interfacial and vdW-gap contributions represented in Fig.~\ref{Fig:TIHGS_Concept}, whereas $C_{\mathrm{TIHGS}}$ used below denotes the BiF\textsubscript{3}/buried-boundary element, with $C_{\mathrm{TIHGS}}^{-1}=C_{\mathrm{BiF_3}}^{-1}+C_Q^{-1}$. The architectural change $C_{\mathrm{d,g}}\rightarrow C_Q$ replaces one finite gate-side response with another whose physical origin, position and electronic character are set by the conversion-generated heterophase. Screening-boundary engineering thus concerns where and how the gate-side electronic response is established, alongside the thickness and permittivity of the dielectric. Supplementary Note~\ref{SN:Screening_Boundary} develops this reduced electrostatic description.

We established the conversion-defined geometry independently through structural and chemical characterization. Cross-sectional transmission electron microscopy (TEM) of fluorinated structures shows an amorphous converted BiF\textsubscript{3} region above residual crystalline Bi\textsubscript{2}Se\textsubscript{3}, with the conversion depth approaching a self-limited value of $\sim$40--43~nm. High-angle annular dark-field scanning transmission electron microscopy (HAADF-STEM) combined with energy-dispersive X-ray spectroscopy (EDS) further resolves the corresponding compositional stratification (Extended Data Fig.~\ref{Fig:TEM_SRIM_EDS}). Atomic force microscopy (AFM), Raman spectroscopy and X-ray photoelectron spectroscopy (XPS) independently support spatially selective chemical conversion rather than additive deposition: for a 5-s treatment, the $\sim$3.3-nm surface-height increase measured by AFM is substantially smaller than the $\sim$13.1-nm converted-layer thickness measured by TEM, consistent with conversion and expansion of the parent Bi\textsubscript{2}Se\textsubscript{3}, while XPS directly reveals Bi--F bonding (Extended Data Fig.~\ref{Fig:AFM_Raman_XPS}). The conversion-depth dependence and its relation to fluorine penetration are discussed in Supplementary Note~\ref{SN:SRIM}.

A functional TIHGS requires two conditions to coexist: the converted surface phase must provide dielectric isolation, while the residual Bi\textsubscript{2}Se\textsubscript{3} beneath it must remain electrically conducting. A pristine Bi\textsubscript{2}Se\textsubscript{3} channel carries current in the mA~$\mu$m$^{-1}$ range, whereas a fully converted channel and a lateral pristine/converted junction are suppressed to the pA~$\mu$m$^{-1}$ range (Fig.~\ref{Fig:Electrical_Realization}a--c). The converted region therefore acts as an electrically isolating barrier rather than simply a resistive modification of Bi\textsubscript{2}Se\textsubscript{3}. Comparable electrical isolation was not reproduced by O\textsubscript{2}-plasma treatment, under which Bi\textsubscript{2}Se\textsubscript{3} remained conductive even after prolonged exposure (Extended Data Fig.~\ref{Fig:Electrical_Controls}a--h).

\begin{figure*}[!]
\centering
\includegraphics[width=\textwidth]{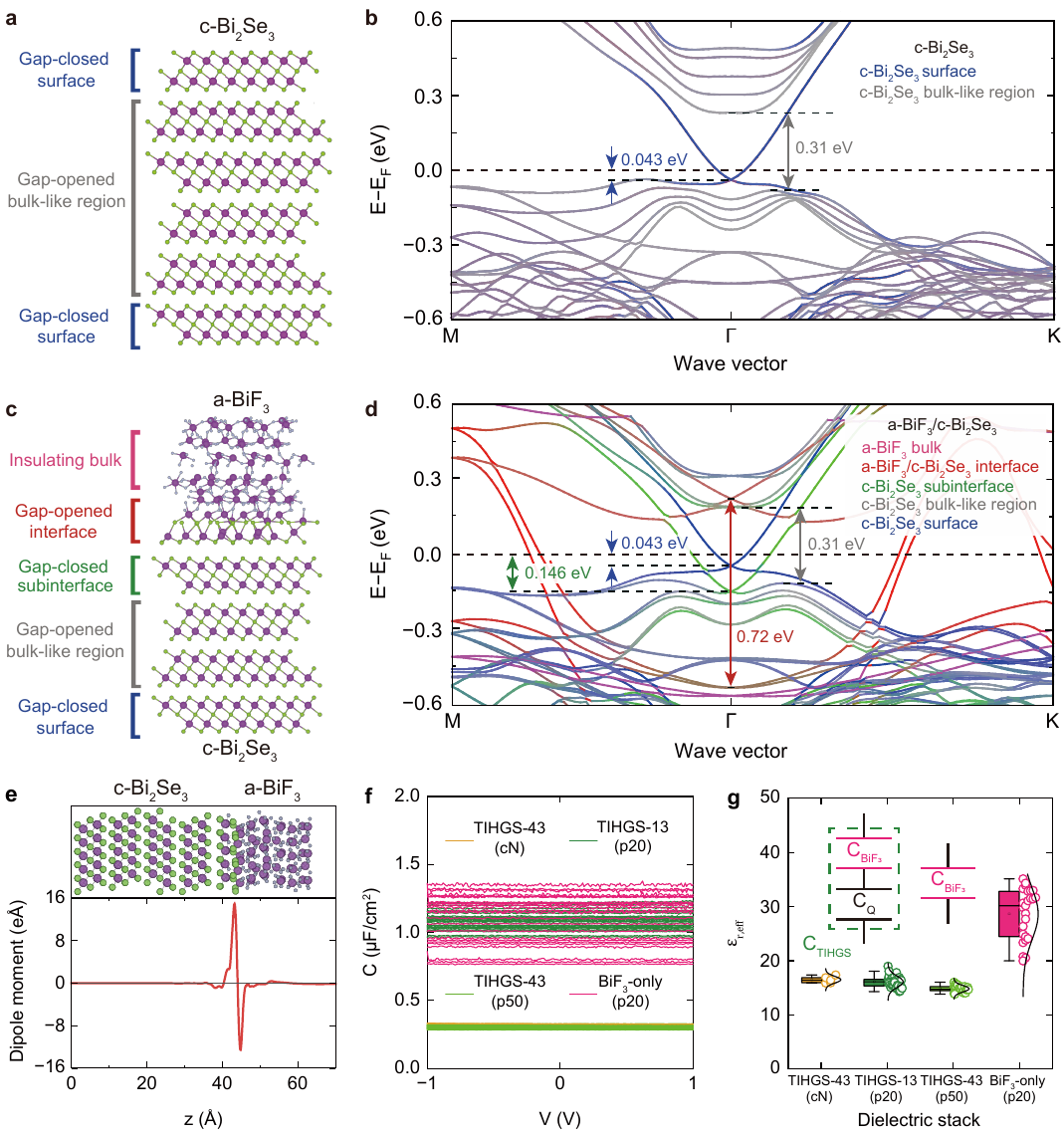}
\caption{\textbf{Electronic reconstruction and electrostatic response of the buried TIHGS boundary.}
\textbf{a}, Six-quintuple-layer (QL) crystalline Bi\textsubscript{2}Se\textsubscript{3} (c-Bi\textsubscript{2}Se\textsubscript{3}) (0001) slab used to model the pristine surface, with gap-closed outer QLs and a gap-opened bulk-like interior.
\textbf{b}, Surface-weighted electronic band structure of \textbf{a}. Bulk-projected states are grey and surface-weighted states blue; the calculated bulk gap is 0.31~eV and the Dirac point lies $\sim$0.043~eV below the Fermi level ($E_{\mathrm{F}}$).
\textbf{c}, Structural model of amorphous BiF\textsubscript{3} (a-BiF\textsubscript{3}) interfaced with four-QL c-Bi\textsubscript{2}Se\textsubscript{3}, showing the immediate heterophase interface, adjacent subinterface QL, bulk-like region, and opposite surface.
\textbf{d}, Position-resolved electronic structure of \textbf{c}. The immediate interface is gap-opened by $\sim$0.72~eV, whereas the adjacent subinterface QL exhibits a reconstructed gap-closed Bi\textsubscript{2}Se\textsubscript{3}-derived Dirac-like feature centred $\sim$0.146~eV below $E_{\mathrm{F}}$ with finite near-$E_{\mathrm{F}}$ spectral weight. The opposite surface remains pristine-like.
\textbf{e}, Calculated out-of-plane dipole-moment profile, showing a response localized near the heterophase interface.
\textbf{f}, Capacitance--voltage ($C$--$V$) characteristics of BiF\textsubscript{3}-only (p20), TIHGS-13 (p20), TIHGS-43 (p50) and TIHGS-43 (c$N$) MIM capacitors ($n=25$, 33, 24 and 5, respectively). TIHGS-13 and TIHGS-43 contain $\sim$13- and $\sim$43-nm converted BiF\textsubscript{3}; p20 and p50 denote nominal starting polycrystalline Bi\textsubscript{2}Se\textsubscript{3} thicknesses, and c$N$ crystalline Bi\textsubscript{2}Se\textsubscript{3} with nominal starting thickness $N$.
\textbf{g}, Distributions of $\varepsilon_{\mathrm{BiF_3}}$ and the BiF\textsubscript{3}-thickness-normalized $\varepsilon_{\mathrm{eff,TIHGS}}$. Inset, series-capacitance representation of $C_{\mathrm{BiF_3}}$ and effective interfacial quantum capacitance $C_Q$. Boxes indicate the interquartile range, with centre lines denoting the median, and individual symbols represent separate capacitors.}
\label{Fig:Buried_Boundary}
\end{figure*}

The complementary requirement of buried conduction is established by conversion-depth-dependent transport. Bi\textsubscript{2}Se\textsubscript{3} flakes thicker than the self-limited conversion depth remain conductive after partial fluorination, whereas thinner flakes become electrically isolated after complete conversion (Fig.~\ref{Fig:Electrical_Realization}d--f). Quantitative transfer-length measurements across a locally converted region further show nearly unchanged contact resistance ($\sim$157 versus $\sim$161~$\Omega$) together with a $\sim$3.5-fold increase in sheet resistance, consistent with an electrically continuous Bi\textsubscript{2}Se\textsubscript{3} pathway persisting beneath the insulating converted region (Extended Data Fig.~\ref{Fig:Electrical_Controls}i--m). These measurements establish the vertically stratified insulating-BiF\textsubscript{3}/conducting-Bi\textsubscript{2}Se\textsubscript{3} configuration required for the buried gate-side screening boundary.

The electrically stratified BiF\textsubscript{3}/Bi\textsubscript{2}Se\textsubscript{3} structure raises a central microscopic question: where does the electronically active boundary reside after surface conversion? Pristine c-Bi\textsubscript{2}Se\textsubscript{3} exhibits a Dirac-like surface-state dispersion within a calculated 0.31-eV bulk gap, with the Dirac-point energy ($E_{\mathrm{D}}$) located $\sim$0.043~eV below the Fermi level ($E_{\mathrm{F}}$) and finite spectral weight extending to $E_{\mathrm{F}}$ (Fig.~\ref{Fig:Buried_Boundary}a,b). Formation of the a-BiF\textsubscript{3}/c-Bi\textsubscript{2}Se\textsubscript{3} heterophase produces a spatially resolved reconstruction. The immediate heterophase interface becomes gap-opened, with an interface-projected gap of $\sim$0.72~eV, whereas the adjacent c-Bi\textsubscript{2}Se\textsubscript{3} subinterface QL develops a reconstructed gap-closed Bi\textsubscript{2}Se\textsubscript{3}-derived Dirac-like feature centred $\sim$0.146~eV below $E_{\mathrm{F}}$ and retaining finite near-$E_{\mathrm{F}}$ spectral weight (Fig.~\ref{Fig:Buried_Boundary}c,d). The opposite surface remains pristine-like. The direct microscopic result is therefore a spatial separation between the gap-opened chemical interface and an electronically active gap-closed subinterface boundary, rather than simple survival of the original Bi\textsubscript{2}Se\textsubscript{3} surface state.

The calculated dipole-moment profile is strongly localized near the a-BiF\textsubscript{3}/c-Bi\textsubscript{2}Se\textsubscript{3} heterophase interface (Fig.~\ref{Fig:Buried_Boundary}e), identifying interfacial charge redistribution accompanying surface conversion. Together, the localized dipole and position-resolved band reconstruction show that conversion reorganizes the Bi\textsubscript{2}Se\textsubscript{3}-derived electronic boundary while creating an insulating chemical interface. The resulting spatial distinction between the chemical conversion front and the gap-closed subinterface state provides the microscopic basis for the buried-boundary picture developed in Supplementary Note~3. Its electronic compressibility is examined independently below through capacitance measurements. Complementary calculations confirm that amorphous BiF\textsubscript{3} remains insulating, with a calculated band gap of 3.94~eV; the crystalline and amorphous BiF\textsubscript{3} electronic structures and calculated dielectric properties are summarized in Extended Data Fig.~\ref{Fig:BiF3_DFT} and Extended Data Table~\ref{tab:bandgap_permittivity}.

Capacitor measurements provide an independent electrostatic test of the buried electronic boundary. Crystalline TIHGS-43 (c$N$) MIM capacitors containing a $\sim$43-nm-thick converted BiF\textsubscript{3} layer exhibit nearly bias-independent capacitance and a mean BiF\textsubscript{3}-thickness-normalized effective relative permittivity $\varepsilon_{\mathrm{eff,TIHGS}}=16.50$ (Fig.~\ref{Fig:Buried_Boundary}f,g). Polycrystalline TIHGS-13 (p20) and TIHGS-43 (p50) similarly yield $\varepsilon_{\mathrm{eff,TIHGS}}\approx15$--16, whereas the fully converted BiF\textsubscript{3}-only reference yields $\varepsilon_{\mathrm{BiF_3}}=28.65$ on average. The lower TIHGS capacitance is represented by
\[
\frac{1}{C_{\mathrm{TIHGS}}}
=
\frac{1}{C_Q}
+
\frac{1}{C_{\mathrm{BiF_3}}},
\]
where the independently measured BiF\textsubscript{3} reference gives effective $C_Q$ values of $\sim$2.52 and $\sim$0.631~$\mu$F~cm$^{-2}$ for TIHGS-13 (p20) and TIHGS-43 (p50), respectively (Supplementary Note~4). Their geometry dependence shows that $C_Q$ is an effective interfacial response rather than a universal material constant. Most importantly, the finite $C_Q$ is a series response that reduces the measured stack capacitance. Combined with the independently calculated near-$E_{\mathrm{F}}$ subinterface spectral weight, this finite electronic response provides mutually consistent microscopic and electrostatic evidence for an electronically compressible buried boundary. The weak dependence of $\varepsilon_{\mathrm{eff,TIHGS}}$ on the nominal starting crystalline Bi\textsubscript{2}Se\textsubscript{3} thickness at a nearly fixed converted BiF\textsubscript{3} thickness further disfavors a dominant bulk contribution (Extended Data Fig.~\ref{Fig:MIM_Ref}f). The broader dielectric-property context is shown in Extended Data Fig.~\ref{Fig:MIM_Ref}g.

\begin{figure*}[!b]
\centering
\includegraphics[width=\textwidth]{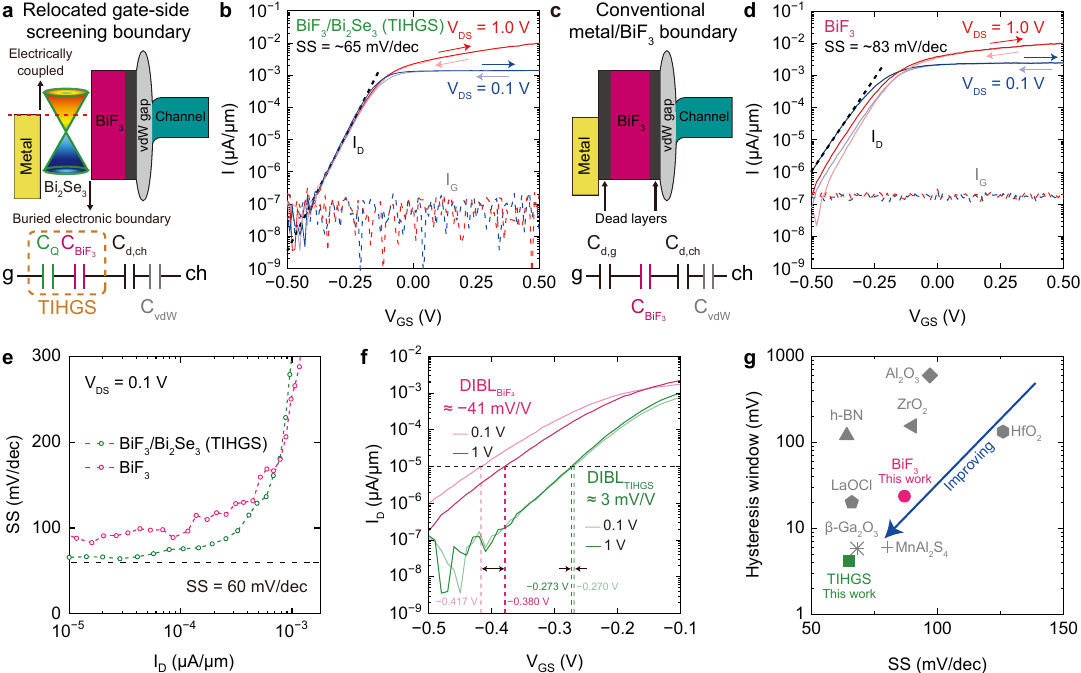}
\caption{\textbf{Enhanced electrostatic control in TIHGS-gated MoS\textsubscript{2} transistors.}
\textbf{a,c}, Schematics and reduced equivalent-capacitance models of TIHGS-gated \textbf{(a)} and BiF\textsubscript{3}-only-gated \textbf{(c)} MoS\textsubscript{2} field-effect transistors (FETs). Capacitance symbols follow the definitions in Fig.~\ref{Fig:TIHGS_Concept}. In \textbf{a}, residual Bi\textsubscript{2}Se\textsubscript{3} remains electrically coupled to the gate electrode, and $C_Q$ represents the effective electronic-compressibility response of the buried electronic boundary in the Bi\textsubscript{2}Se\textsubscript{3} subinterface region. The dashed circuit box identifies the TIHGS element comprising $C_Q$ and $C_{\mathrm{BiF_3}}$ in series. In \textbf{c}, $C_{\mathrm{d,g}}$ represents the conventional metal/BiF\textsubscript{3} interfacial response. Both models retain the BiF\textsubscript{3} dielectric and channel-side contributions; the regions labelled ``Dead layers'' schematically represent effective interfacial responses. The physical BiF\textsubscript{3} thicknesses are closely matched at $\sim$43~nm in both devices.
\textbf{b,d}, Forward and reverse transfer characteristics of representative TIHGS-gated \textbf{(b)} and BiF\textsubscript{3}-gated \textbf{(d)} devices. The TIHGS-gated device exhibits a minimum subthreshold swing ($SS$) of $\sim$65~mV~dec$^{-1}$ and negligible hysteresis, compared with $\sim$83~mV~dec$^{-1}$ and larger hysteresis for the BiF\textsubscript{3}-gated control. Coloured dashed traces denote gate-leakage current.
\textbf{e}, Local $SS$ as a function of width-normalized drain current, extracted from the forward transfer characteristics at $V_{\mathrm{DS}}=0.1$~V. The dashed line marks the room-temperature thermionic limit of 60~mV~dec$^{-1}$.
\textbf{f}, Constant-current extraction of drain-induced barrier lowering ($DIBL$) from the forward subthreshold transfer characteristics. Magenta and green traces denote the BiF\textsubscript{3}-only and TIHGS-gated devices, respectively; light and dark traces correspond to $V_{\mathrm{DS}}=0.1$ and 1.0~V. The horizontal dashed line marks $I_{\mathrm{D}}=10$~pA~$\mu$m$^{-1}$. Vertical dashed lines identify the corresponding gate voltages, and arrows indicate their separation. The $DIBL$ values are approximately $+3$ and $-41$~mV~V$^{-1}$ for the TIHGS- and BiF\textsubscript{3}-only-gated devices, respectively.
\textbf{g}, Comparison of $SS$ and hysteresis window for the TIHGS- and BiF\textsubscript{3}-gated devices from this work and representative reported MoS\textsubscript{2} FETs employing different gate dielectrics~\cite{hu2026ultrahigh,xu2022few,chang2021ald,fu2025low,wen2016effects,zou2019improved,shen2025mos2}. Literature values correspond to devices with different geometries and were obtained using different gate-voltage ranges, sweep conditions, and extraction protocols; they are therefore shown for contextual benchmarking rather than direct device-to-device ranking.}
\label{Fig:FET}
\end{figure*}

For this boundary architecture to function as a gate stack, the converted BiF\textsubscript{3} phase must also retain sufficient electrical isolation. For positive capacitor biases up to approximately 0.5~V, the TIHGS MIM current density remains below $10^{-2}$~A~cm$^{-2}$. The transistor measurements separately resolve the drain current above the gate-leakage level during turn-on, whereas the deepest off-state currents approach the leakage and measurement-noise floor. At higher fields, the MIM leakage characteristics evolve from field-assisted emission towards Fowler--Nordheim tunnelling. The localized dipole in Fig.~\ref{Fig:Buried_Boundary}e provides microscopic evidence of interfacial charge redistribution, whereas Extended Data Fig.~\ref{Fig:TIHGS_Band} distinguishes the measured leakage regimes and the vacuum-referenced, contact-equilibrated and high-field band-alignment configurations. The complete transport analysis and its limitations are discussed in Supplementary Note~\ref{SN:Emission_Tunnelling}. Frequency-dependent capacitor measurements further show modest dispersion and low loss from 1~kHz to 1~MHz (Extended Data Fig.~\ref{Fig:MIM_Ref}c--e), and WSe\textsubscript{2}/TIHGS MISCAPs reproduce a stack-level $\varepsilon_{\mathrm{eff,TIHGS}}\approx16$ after semiconductor integration (Extended Data Fig.~\ref{Fig:MISCAP}, Extended Data Table~\ref{tab:dielectric_stack}, and Supplementary Notes~\ref{SN:Flat-Band_Screening} and \ref{SN:Interface-Trap}).

We next evaluated the device-level effect of the gate-side boundary using MoS\textsubscript{2} FETs incorporating either a TIHGS or a fully converted BiF\textsubscript{3}-only dielectric. The representative TIHGS-gated and BiF\textsubscript{3}-gated devices had closely matched insulating BiF\textsubscript{3} thicknesses of approximately 43 and 42.8~nm, respectively, and channel lengths of 8.56 and 9.30~$\mu$m. Both devices retained a common BiF\textsubscript{3}/MoS\textsubscript{2} channel-side material interface while differing in their gate-side boundary condition (Fig.~\ref{Fig:FET}a,c). The TIHGS-gated transistor exhibits an on--off current ratio exceeding $10^{6}$, a minimum room-temperature subthreshold swing ($SS$) of $\sim$65~mV~dec$^{-1}$ and negligible hysteresis, compared with $\sim10^{5}$, $\sim83$~mV~dec$^{-1}$ and larger hysteresis for the BiF\textsubscript{3}-gated control (Fig.~\ref{Fig:FET}b--e). Given the closely matched dielectric thicknesses and independently measured lower nominal TIHGS stack capacitance, the improved switching characteristics cannot be attributed simply to a thinner or higher-capacitance dielectric.

At a common width-normalized drain current of $I_{\mathrm{D}}=10$~pA~$\mu$m$^{-1}$, constant-current analysis yields signed $DIBL$ values of approximately $+3$ and $-41$~mV~V$^{-1}$ for the representative TIHGS- and BiF\textsubscript{3}-only-gated devices, respectively (Fig.~\ref{Fig:FET}f). The TIHGS therefore exhibits a substantially smaller magnitude of subthreshold drain-bias dependence.

An air-oxidized BiO\textsubscript{x}/Bi\textsubscript{2}Se\textsubscript{3} control exhibits a minimum $SS$ of $\sim75$~mV~dec$^{-1}$ and $DIBL\approx-14$~mV~V$^{-1}$ at the higher current criterion of $I_{\mathrm{D}}=100$~pA~$\mu$m$^{-1}$ (Extended Data Fig.~\ref{Fig:FET_Controls}g). The smaller extracted $|DIBL|$ in both Bi\textsubscript{2}Se\textsubscript{3}-containing stacks at their respective current criteria is qualitatively consistent with a contribution from the buried gate-side interface, although their dielectric structures also differ. Device-to-device measurements further show that the TIHGS combines low $SS$ with substantially lower gate leakage than the oxide control (Extended Data Fig.~\ref{Fig:FET_Controls}h).

Contextual literature benchmarking also places the representative TIHGS device in the low-$SS$/low-hysteresis region (Fig.~\ref{Fig:FET}g)~\cite{hu2026ultrahigh,xu2022few,chang2021ald,fu2025low,wen2016effects,zou2019improved,shen2025mos2}. The matched TIHGS/BiF\textsubscript{3}-only comparison links improved switching and weaker subthreshold drain-bias dependence to a conversion-defined gate architecture with lower nominal capacitance. Independently, finite effective $C_Q$ and the calculated subinterface states support an electronically compressible buried-boundary picture. These complementary results identify the gate-side electronic boundary as a design degree of freedom connecting interfacial electronic reconstruction with transistor electrostatics.

\section{Conclusions}
\label{sec3}

Surface conversion of Bi\textsubscript{2}Se\textsubscript{3} provides a materials-based route to forming an insulating BiF\textsubscript{3} dielectric while reconstructing a buried gate-side electronic boundary. Position-resolved calculations distinguish the gap-opened chemical interface from an adjacent gap-closed subinterface state and reveal localized interfacial charge redistribution. The calculated near-$E_{\mathrm{F}}$ spectral weight and independently measured finite series response together support an electronically compressible buried boundary. In MoS\textsubscript{2} transistors with closely matched BiF\textsubscript{3} thicknesses and a common channel-side material interface, this architecture combines near-thermionic-limit switching, negligible hysteresis, and weak subthreshold drain-bias dependence despite its lower nominal capacitance. Our work establishes a gate-stack design strategy that distinguishes the chemical interface, electronic screening boundary, and dielectric capacitance. Engineering the identity and location of the screening boundary therefore complements dielectric scaling and channel-side interface optimization.

\section{Methods}
\label{sec4}

\subsection{First-Principles Calculations}
\label{M1}

Density functional theory (DFT) calculations were performed using the Vienna \textit{Ab initio} Simulation Package (VASP)~\cite{kresse1996efficient}. All DFT calculations used the Perdew--Burke--Ernzerhof (PBE) functional within the generalized gradient approximation (GGA)~\cite{perdew1996generalized}, projector-augmented-wave (PAW) potentials~\cite{blochl1994projector}, and spin--orbit coupling (SOC); this framework is hereafter denoted PBE+SOC. A plane-wave energy cutoff of 520~eV was used. Structures used for the DFT analyses were relaxed until the residual Hellmann--Feynman force on each atom was below 0.01~eV~\AA$^{-1}$. Long-range dispersion interactions were included using the DFT-D3 method with Becke--Johnson damping~\cite{grimme2011effect}. The first-principles calculations presented in Fig.~\ref{Fig:Buried_Boundary}a--e and Extended Data Fig.~\ref{Fig:BiF3_DFT}, together with the calculated properties summarized in Extended Data Table~\ref{tab:bandgap_permittivity}, were obtained within this DFT framework. Dielectric responses were evaluated using density-functional perturbation theory (DFPT)~\cite{gonze1997dynamical}.

Amorphous BiF\textsubscript{3}, Bi\textsubscript{2}Se\textsubscript{3}, and Bi\textsubscript{2}O\textsubscript{3} structures were generated using melt--quench molecular-dynamics simulations with the SevenNet-Omni-i12 machine-learning interatomic potential~\cite{kim2026optimizing}. The resulting amorphous structures were subsequently relaxed within the DFT framework described above before electronic-structure and dielectric calculations.

For bulk Bi\textsubscript{2}Se\textsubscript{3}, Bi\textsubscript{2}O\textsubscript{3}, and BiF\textsubscript{3} structures, the Brillouin zone was sampled using $k$-point meshes corresponding to a reciprocal-space density of 100 $k$-points per \AA$^{-3}$. The pristine Bi\textsubscript{2}Se\textsubscript{3} slab and the a-BiF\textsubscript{3}/c-Bi\textsubscript{2}Se\textsubscript{3} heterophase slab were sampled using $5 \times 5 \times 1$ $k$-point meshes.

For the pristine Bi\textsubscript{2}Se\textsubscript{3} surface calculation, a six-quintuple-layer (QL) crystalline Bi\textsubscript{2}Se\textsubscript{3} slab terminated along the (0001) direction was used (Fig.~\ref{Fig:Buried_Boundary}a,b). To minimize direct interactions between adjacent periodic images, all slab structures were constructed with a vacuum spacing greater than 15~\AA.

For the heterophase calculation, an amorphous BiF\textsubscript{3} region was interfaced with a four-QL crystalline Bi\textsubscript{2}Se\textsubscript{3} slab to model the buried BiF\textsubscript{3}/Bi\textsubscript{2}Se\textsubscript{3} boundary. The QL-resolved electronic structures were obtained directly from the VASP calculations by projecting the electronic states onto individual Bi\textsubscript{2}Se\textsubscript{3} QLs at different distances from the heterophase boundary. We compare the subinterface QL, corresponding to the first Bi\textsubscript{2}Se\textsubscript{3} QL directly beneath the BiF\textsubscript{3}/Bi\textsubscript{2}Se\textsubscript{3} interface, with the surface QL at the opposite side of the four-QL slab, corresponding to the fourth Bi\textsubscript{2}Se\textsubscript{3} QL counted from the heterophase boundary (Fig.~\ref{Fig:Buried_Boundary}c,d).

\subsection{SRIM Simulations}
\label{M2}

Stopping and range of ions in matter (SRIM) simulations were performed using the SRIM-2013 package~\cite{ziegler2010srim}. We used a 50-nm-thick stoichiometric Bi\textsubscript{2}Se\textsubscript{3} layer as a representative target geometry. The displacement energies of Bi and Se were set to 30~eV based on the values reported by Konobeyev \textit{et al.}~\cite{konobeyev2017evaluation}. Because the ion-energy distribution incident on the sample surface was not measured experimentally, F\textsuperscript{+} incident energies of 0.5, 1, 2.5, and 5~keV were examined as a sensitivity range for the accessible ballistic penetration length scale. The 5-keV calculation was deliberately included as a high-energy upper-bound case and should not be interpreted as the dominant fluorine-ion energy under the experimental plasma conditions. For each incident energy, we simulated 10,000 ions to obtain statistically stable fluorine depth distributions. To compare incident energies, the resulting fluorine depth profiles were area-normalized and smoothed with a Gaussian filter.

The SRIM calculations describe ballistic ion transport only and do not include plasma chemistry, transport of neutral reactive species, fluorine diffusion, sputter erosion, dynamic changes in target composition or density, surface charging, plasma-sheath evolution, or formation of BiF\textsubscript{3}. The calculations therefore provide a ballistic length-scale constraint on fluorine penetration rather than a quantitative description of fluorination kinetics, as discussed in Supplementary Note~\ref{SN:SRIM}.

\subsection{Synthesis of Polycrystalline Bi\textsubscript{2}Se\textsubscript{3} Films}
\label{M3}

Polycrystalline Bi\textsubscript{2}Se\textsubscript{3} films were synthesized by solution deposition followed by thermal selenization. Precursor solutions with concentrations of 5 and 10~wt\% were prepared by dissolving bis\{tris[\textit{N}-ethoxy-2,2-dimethylpropanamidato]bismuth(III)\} ([Bi(edpa)\textsubscript{3}]\textsubscript{2}) in tetrahydrofuran (THF) and stirring for 60~min to obtain homogeneous solutions. The solutions were spin-coated onto 300-nm-thick SiO\textsubscript{2}/Si substrates that had been rendered hydrophilic by ultraviolet--ozone treatment for 600~s. Spin coating was performed at 1,000~rpm for 10~s followed by 2,000~rpm for 30~s. The resulting precursor layers were immediately annealed at 100~$^\circ$C for 1~min to remove residual solvent. The 5 and 10~wt\% precursor concentrations produced Bi\textsubscript{2}Se\textsubscript{3} films with nominal thicknesses of approximately 20 and 50~nm, respectively, after selenization.

For selenization, the precursor-coated substrates were placed in a quartz-tube furnace with 0.2~g of Se powder (99.99\%, Sigma-Aldrich) positioned upstream. The furnace was purged and maintained under 99.9999\% Ar at a flow rate of 1,000 standard cubic centimetres per minute (sccm) and a pressure of 3~Torr. The temperature was increased to 300~$^\circ$C at a rate of 15~$^\circ$C~min$^{-1}$ and maintained for 60~min to induce precursor decomposition and selenization, yielding large-area polycrystalline Bi\textsubscript{2}Se\textsubscript{3} films. After selenization, the furnace was turned off and allowed to cool naturally to room temperature under the Ar atmosphere.

\subsection{Spatially Selective Fluorination of Bi\textsubscript{2}Se\textsubscript{3}}
\label{M4}

Crystalline Bi\textsubscript{2}Se\textsubscript{3} flakes used for spatially selective fluorination and transport measurements were mechanically exfoliated from bulk single crystals and transferred onto highly doped Si substrates capped with 285~nm SiO\textsubscript{2}. Selected regions of the Bi\textsubscript{2}Se\textsubscript{3} surface were exposed using electron-beam lithography (Vistec EBPG 5000), while the regions intended to remain pristine were protected with polymethyl methacrylate (PMMA).

Surface conversion was performed in an Oxford Instruments PlasmaLab 80 Plus system using an SF\textsubscript{6}/Ar plasma with SF\textsubscript{6} and Ar flow rates of 10 and 5~sccm, respectively, a chamber pressure of 50~mTorr, and a radio-frequency (RF) power of 50~W. The converted-layer thickness increased with plasma-treatment time and approached a self-limited thickness of $\sim$40--43~nm after $\sim$20--25~s of treatment (Extended Data Fig.~\ref{Fig:TEM_SRIM_EDS}). A 5-s treatment produced a $\sim$13.1-nm-thick converted surface layer under the conditions used for the thickness-calibration experiment, whereas a 25-s treatment produced a $\sim$42.5--43-nm-thick converted layer. Regions protected by PMMA remained unexposed to the SF\textsubscript{6} plasma.

For the plasma-control experiments in Extended Data Fig.~\ref{Fig:Electrical_Controls}a--h, Bi\textsubscript{2}Se\textsubscript{3} was exposed to O\textsubscript{2} plasma at an O\textsubscript{2} flow rate of 30~sccm, a chamber pressure of 30~mTorr, and an RF power of 50~W for either 10 or 60~s.

For the lateral-transport and transfer-length-method structures in Fig.~\ref{Fig:Electrical_Realization} and Extended Data Fig.~\ref{Fig:Electrical_Controls}, we patterned metal electrodes using electron-beam lithography and deposited 5~nm Ti and then 50~nm Au via electron-beam evaporation (MASA IM-9912), followed by lift-off. Hexagonal boron nitride (h-BN) flakes were deterministically transferred onto the devices as protective encapsulation using a dry-release transfer process based on polypropylene carbonate (PPC) and polydimethylsiloxane (PDMS).

\subsection{Fabrication of BiF\textsubscript{3} and TIHGS MIM Capacitors}
\label{M5}

Polycrystalline Bi\textsubscript{2}Se\textsubscript{3} films with nominal starting thicknesses of approximately 20 and 50~nm were used to fabricate the fully converted BiF\textsubscript{3} reference and partially converted topological-insulator heterophase gate stack (TIHGS) capacitor structures. The films were transferred onto substrates coated with 5~nm Ti and 50~nm Au on 285-nm-thick SiO\textsubscript{2}/Si using a wet-transfer process.

Fluorination was carried out using the SF\textsubscript{6}/Ar plasma conditions described above. Three polycrystalline conversion configurations were used. A nominally 20-nm-thick starting Bi\textsubscript{2}Se\textsubscript{3} film treated for 5~s produced a $\sim$13-nm-thick converted BiF\textsubscript{3} layer while retaining underlying Bi\textsubscript{2}Se\textsubscript{3} and was denoted TIHGS-13 (p20). A nominally 50-nm-thick starting film treated for 25~s produced a $\sim$43-nm-thick converted BiF\textsubscript{3} layer while retaining underlying Bi\textsubscript{2}Se\textsubscript{3} and was denoted TIHGS-43 (p50). In contrast, a nominally 20-nm-thick starting film treated for 25~s was fully converted and used as the BiF\textsubscript{3}-only reference [BiF\textsubscript{3}-only (p20)]. Atomic force microscopy independently determined the post-conversion physical thickness of this fully converted BiF\textsubscript{3} layer to be $\sim$23~nm, and this measured thickness was used for extraction of $\varepsilon_{\mathrm{BiF_3}}$. For the partially converted TIHGS structures, the converted BiF\textsubscript{3} thickness and nominal starting Bi\textsubscript{2}Se\textsubscript{3} thickness are reported independently because fluorination is accompanied by volume expansion and the residual Bi\textsubscript{2}Se\textsubscript{3} thickness cannot be obtained by simple subtraction.

For capacitance measurements of the polycrystalline structures, a 30-nm-thick Al\textsubscript{2}O\textsubscript{3} layer was deposited after fluorination by atomic layer deposition (ALD) at 300~$^\circ$C using trimethylaluminum (TMA) and H\textsubscript{2}O as the Al and O precursors, respectively. Each ALD cycle consisted of a 0.5-s TMA pulse, a 2-s Ar purge, a 0.5-s H\textsubscript{2}O pulse, and a 2-s Ar purge, and a total of 270 cycles was used to obtain the nominal 30-nm thickness. Independently fabricated 30-nm Al\textsubscript{2}O\textsubscript{3} reference capacitors prepared under the same ALD conditions were used to determine the Al\textsubscript{2}O\textsubscript{3} dielectric response and to de-embed its series contribution from the BiF\textsubscript{3}-only and TIHGS structures. Top electrodes were defined using a shadow mask, followed by electron-beam evaporation of 15~nm Pt, 150~nm Ag, and 20~nm Au, giving an Au/Ag/Pt top-electrode stack.

Crystalline-Bi\textsubscript{2}Se\textsubscript{3}-based TIHGS MIM capacitors were fabricated separately from mechanically exfoliated Bi\textsubscript{2}Se\textsubscript{3} flakes with nominal starting thicknesses of 49, 65, 91, 128, or 179~nm. The Bi\textsubscript{2}Se\textsubscript{3} flakes were first transferred onto bottom electrodes comprising 5~nm Ti and 50~nm Au. These samples were fluorinated under conditions producing a $\sim$43-nm-thick converted BiF\textsubscript{3} layer and are denoted TIHGS-43 (c$N$), where $N$ is the nominal starting crystalline-Bi\textsubscript{2}Se\textsubscript{3} thickness in nanometres. Following fluorination, 100-nm-thick Au top electrodes were deposited directly onto the converted BiF\textsubscript{3} surface without an intervening Ti adhesion layer, yielding Au/BiF\textsubscript{3}/Bi\textsubscript{2}Se\textsubscript{3}/Au/Ti MIM structures.

\subsection{Fabrication of WSe\textsubscript{2}/TIHGS MISCAPs}
\label{M6}

Metal--insulator--semiconductor capacitors (MISCAPs) were fabricated using crystalline-Bi\textsubscript{2}Se\textsubscript{3}-based TIHGS structures containing a $\sim$43-nm-thick converted BiF\textsubscript{3} layer. Bi\textsubscript{2}Se\textsubscript{3} flakes were first transferred onto bottom electrodes comprising 5~nm Ti and 50~nm Au. A 37.37-nm-thick WSe\textsubscript{2} flake, as determined by atomic force microscopy, was deterministically transferred onto the BiF\textsubscript{3} surface using a PPC/PDMS-based dry-release transfer process, thereby forming the semiconductor/dielectric interface. Following lithography, top electrodes comprising 15~nm Pt, 150~nm Ag, and 20~nm Au were deposited by electron-beam evaporation. The completed MISCAP therefore had an Au/Ag/Pt/WSe\textsubscript{2}/BiF\textsubscript{3}/Bi\textsubscript{2}Se\textsubscript{3}/Au/Ti vertical stack. The underlying Bi\textsubscript{2}Se\textsubscript{3} remained electrically coupled to the bottom Au/Ti electrode, while WSe\textsubscript{2} formed the semiconductor side of the MISCAP.

\subsection{Fabrication of MoS\textsubscript{2} Field-Effect Transistors}
\label{M7}

MoS\textsubscript{2} FETs were fabricated using three gate-stack configurations: a TIHGS comprising partially converted BiF\textsubscript{3}/Bi\textsubscript{2}Se\textsubscript{3}, a fully converted BiF\textsubscript{3}-only dielectric, and an air-oxidized BiO\textsubscript{x}/Bi\textsubscript{2}Se\textsubscript{3} control stack.

For the TIHGS-gated devices, Bi\textsubscript{2}Se\textsubscript{3} flakes thicker than the self-limited fluorination depth were partially converted so that an insulating BiF\textsubscript{3} region directly faced the MoS\textsubscript{2} channel while electrically conducting Bi\textsubscript{2}Se\textsubscript{3} remained beneath the converted region and was electrically coupled to the metal gate. For the BiF\textsubscript{3}-gated control devices, Bi\textsubscript{2}Se\textsubscript{3} flakes thinner than the self-limited conversion depth were fully fluorinated, producing a BiF\textsubscript{3}-only dielectric in direct contact with the metal gate. Thus, the TIHGS-gated and BiF\textsubscript{3}-gated devices shared a common BiF\textsubscript{3}/MoS\textsubscript{2} channel-side interface but differed in their gate-side electrostatic boundary.

Mechanically exfoliated MoS\textsubscript{2} flakes were assembled onto the respective gate-stack structures as follows. Bi\textsubscript{2}Se\textsubscript{3} flakes were first transferred onto bottom electrodes comprising 5~nm Ti and 50~nm Au and were then partially or fully fluorinated under the conditions described above to form the TIHGS or BiF\textsubscript{3}-only gate stack, respectively. MoS\textsubscript{2} flakes approximately 20~nm thick and 30~$\mu$m in lateral dimension were subsequently transferred onto the BiF\textsubscript{3} surface using a PPC/PDMS-based dry-release transfer process. Source and drain electrodes were patterned by electron-beam lithography and formed by electron-beam evaporation of 5~nm Ti followed by 50~nm Au, with Ti directly contacting MoS\textsubscript{2}.

For the chemically derived interface control, Bi\textsubscript{2}Se\textsubscript{3} was exposed to ambient air for 2~days to form an oxidized BiO\textsubscript{x} surface layer. The corresponding oxide thickness was estimated to be $\sim$1.94~nm based on the previously reported oxidation behaviour of Bi\textsubscript{2}Se\textsubscript{3}~\cite{kong2011rapid}. MoS\textsubscript{2} was subsequently integrated onto the BiO\textsubscript{x}/Bi\textsubscript{2}Se\textsubscript{3} stack using the same device-assembly procedure, and source/drain electrodes were fabricated as described above.

For the representative devices compared in Fig.~\ref{Fig:FET}, the insulating BiF\textsubscript{3} thicknesses were closely matched. For the BiF\textsubscript{3}-gated control, AFM measured a Bi\textsubscript{2}Se\textsubscript{3} thickness of 37.2~nm before fluorination and a post-conversion BiF\textsubscript{3} thickness of 42.8~nm after complete fluorination. The TIHGS-gated device contained an approximately 43-nm-thick converted BiF\textsubscript{3} region while retaining electrically conducting Bi\textsubscript{2}Se\textsubscript{3} underneath. Thus, the two devices had closely matched insulating-layer thicknesses while retaining distinct gate-side screening boundaries. The MoS\textsubscript{2} channel dimensions (length $\times$ width) were 8.56~$\mu$m $\times$ 2.93~$\mu$m for the TIHGS-gated device and 9.30~$\mu$m $\times$ 2.12~$\mu$m for the BiF\textsubscript{3}-gated control.

\subsection{Structural and Chemical Characterization}
\label{M8}

Optical microscopy was performed using Olympus BX60 and MX63L microscopes. Atomic force microscopy (AFM) in Extended Data Fig.~\ref{Fig:AFM_Raman_XPS}a,b was performed using a Bruker Dimension Icon in PeakForce Tapping mode with a ScanAsyst-Air cantilever having a nominal tip radius of $\sim$2~nm. AFM was used to determine flake thicknesses, converted-layer-associated surface-height changes, and the WSe\textsubscript{2} thickness used in the MISCAP analysis.

Cross-sectional transmission electron microscopy (TEM) was performed using a JEOL JEM-2200FS. Cross-sectional specimens were prepared by focused-ion-beam milling using a JEOL JIB-4700F. We used high-angle annular dark-field scanning transmission electron microscopy (HAADF-STEM) and energy-dispersive X-ray spectroscopy (EDS) to examine the spatial distributions of Bi, Se, O, Ti, and Au across the fluorinated heterostructures.

Raman spectroscopy in Extended Data Fig.~\ref{Fig:AFM_Raman_XPS}c,d was performed using a WITec Alpha 300 RA+ system with a continuous-wave excitation laser with a wavelength of 532 nm and a power of 10 $\mu$W. Raman mapping was used to compare pristine and plasma-treated regions and to monitor the characteristic Bi\textsubscript{2}Se\textsubscript{3} phonon modes.

X-ray photoelectron spectroscopy (XPS) in Extended Data Fig.~\ref{Fig:AFM_Raman_XPS}e--h was performed using a Kratos Axis Ultra system equipped with a monochromatic Al K$\alpha$ X-ray source with an energy of 1486.96 eV. The analysed area was $\sim0.3\times0.7$~mm$^{2}$. We acquired high-resolution spectra using a pass energy of 20~eV, an energy step of 0.1~eV, and an X-ray spot size of $\sim$200~$\mu$m. Because the finite escape depth of the emitted photoelectrons limits the information depth of XPS, the measured spectra predominantly probe the near-surface chemical composition. This surface sensitivity enables characterization of the chemical changes induced by SF\textsubscript{6} fluorination.

Optical microscopy, AFM, and Raman measurements were performed under ambient laboratory conditions unless otherwise noted. TEM/STEM and XPS measurements were performed under the vacuum conditions intrinsic to the corresponding instruments.

\subsection{Electrical Measurements and Parameter Extraction}
\label{M9}

Electrical measurements were performed at room temperature using a Keithley 2636B source-measure unit for current--voltage measurements and a Keysight E4980A inductance--capacitance--resistance (LCR) meter for capacitance and conductance measurements. Devices were measured using electrical probes or wire-bonded sample holders, as appropriate. Transistor characteristics were measured in the dark to suppress photoinduced current. All measurements were performed in ambient air.

For the transfer-length-method (TLM) analysis in Extended Data Fig.~\ref{Fig:Electrical_Controls}i--m, the total two-terminal resistance was obtained from the linear low-bias $I$--$V$ response and fitted as
\[
R_{\mathrm{T}}
=
2R_{\mathrm{C}}
+
R_{\mathrm{Sh}}\frac{L_{\mathrm{X}}}{W},
\]
where $R_{\mathrm{C}}$ is the contact resistance per contact, $R_{\mathrm{Sh}}$ is the channel sheet resistance, $L_{\mathrm{X}}$ is the electrode spacing, and $W$ is the channel width. $R_{\mathrm{C}}$ was obtained from one half of the fitted zero-length intercept, and $R_{\mathrm{Sh}}$ was obtained from the fitted slope after accounting for the measured channel width.

For the MIM capacitors in Fig.~\ref{Fig:Buried_Boundary}f,g, capacitance--voltage measurements were performed at 1~MHz using an alternating-current (AC) excitation amplitude of 0.05~V. All capacitances reported in the main text were normalized by the active capacitor area. For the Al\textsubscript{2}O\textsubscript{3}-capped polycrystalline structures, the independently measured Al\textsubscript{2}O\textsubscript{3} capacitance was de-embedded according to
\[
\frac{1}{C_{\mathrm{total}}}
=
\frac{1}{C_{\mathrm{Al_2O_3}}}
+
\frac{1}{C_X},
\]
where $C_X=C_{\mathrm{BiF_3}}$ for the fully converted BiF\textsubscript{3}-only structure and $C_X=C_{\mathrm{TIHGS}}$ for the partially converted TIHGS structure. The BiF\textsubscript{3} relative permittivity was calculated using
\[
C_{\mathrm{BiF_3}}
=
\frac{\varepsilon_{0}\varepsilon_{\mathrm{BiF_3}}}
{t_{\mathrm{BiF_3}}},
\]
where $t_{\mathrm{BiF_3}}$ is the independently determined physical thickness of the BiF\textsubscript{3} layer. For the BiF\textsubscript{3}-only (p20) reference, AFM yielded $t_{\mathrm{BiF_3}}\approx23$~nm, which was used to extract a mean $\varepsilon_{\mathrm{BiF_3}}$ of 28.65 and a median value of 30.16. For TIHGS structures, the BiF\textsubscript{3}-thickness-normalized effective relative permittivity was defined as
\[
\varepsilon_{\mathrm{eff,TIHGS}}
=
\frac{C_{\mathrm{TIHGS}}t_{\mathrm{BiF_3}}}{\varepsilon_{0}}.
\]
The electrically conducting residual Bi\textsubscript{2}Se\textsubscript{3} was not included in the geometric dielectric thickness. The effective interfacial quantum capacitance was obtained using
\[
\frac{1}{C_{\mathrm{TIHGS}}}
=
\frac{1}{C_{\mathrm{BiF_3}}}
+
\frac{1}{C_Q}.
\]
Accordingly, $C_Q$ is treated as an effective interfacial electronic-compressibility response rather than as an intrinsic dielectric capacitance or a universal material constant.

Frequency-dependent capacitance and loss measurements shown in Extended Data Fig.~\ref{Fig:MIM_Ref}c--e were performed from 1~kHz to 1~MHz using a direct-current (DC) bias of 0.1~V and an AC excitation amplitude of 0.05~V. The current-density--voltage measurement and corresponding field-dependent transport analyses shown in Extended Data Fig.~\ref{Fig:TIHGS_Band}a--c were obtained using the Keithley 2636B. For the TIHGS MIM leakage measurements, the bias voltage was applied to the top Au electrode in direct contact with BiF\textsubscript{3}, while the bottom Au/Ti electrode electrically coupled to the residual Bi\textsubscript{2}Se\textsubscript{3} was grounded. Under positive top-electrode bias ($V>0$), electron injection therefore occurs from the residual Bi\textsubscript{2}Se\textsubscript{3} side across the buried BiF\textsubscript{3}/Bi\textsubscript{2}Se\textsubscript{3} interface into BiF\textsubscript{3}. For field-dependent leakage analysis, the electric field was calculated as $E=|V|/t_{\mathrm{BiF_3}}$, using only the physical thickness of the insulating BiF\textsubscript{3} layer.

For the WSe\textsubscript{2}/TIHGS MISCAP characterized in Extended Data Fig.~\ref{Fig:MISCAP}, capacitance--voltage and parallel-conductance characteristics were measured from 10~kHz to 1~MHz over the gate-voltage range $-1~\mathrm{V}\leq V_{\mathrm{G}}\leq1~\mathrm{V}$ using an AC excitation amplitude of 0.05~V. The flat-band condition was extracted from the 10-kHz capacitance characteristic because the accumulation-to-depletion transition remained within the accessible low-leakage gate-voltage range at this frequency. The flat-band voltage was identified from the zero crossing of $\mathrm{d}^{2}C/\mathrm{d}V_{\mathrm{G}}^{2}$ associated with the principal accumulation--depletion inflection. The graphical consistency analysis used to extract $\varepsilon_{\mathrm{eff,TIHGS}}$, together with the subsequent extraction of $C_{\mathrm{TIHGS}}$, effective semiconductor screening length $L_{\mathrm{D,eff}}$, and apparent majority-carrier concentration $N_{\mathrm{app}}$, is described in Supplementary Note~\ref{SN:Flat-Band_Screening}. The effective electrically active trap density, $D_{\mathrm{it,eff}}$, was obtained using the 10-kHz/1-MHz high--low-frequency capacitance method described in Supplementary Note~\ref{SN:Interface-Trap}. The extracted electrostatic parameters are summarized in Extended Data Table~\ref{tab:dielectric_stack}. Because the 10-kHz characteristic is not a true quasi-static limit and an independently resolved frequency-dependent TIHGS capacitance was not available, $D_{\mathrm{it,eff}}$ is interpreted as an effective electrically active trap response sampled within the experimental frequency window rather than as a unique microscopic density of states exclusively at the WSe\textsubscript{2}/BiF\textsubscript{3} interface.

For the MoS\textsubscript{2} field-effect transistors (FETs) characterized in Fig.~\ref{Fig:FET} and Extended Data Fig.~\ref{Fig:FET_Controls}, output characteristics were measured over $0\leq V_{\mathrm{DS}}\leq1.0$~V, where $V_{\mathrm{DS}}$ is the drain--source voltage, at gate--source voltages $-0.3~\mathrm{V}\leq V_{\mathrm{GS}}\leq0.5$~V. Transfer characteristics were measured over $-0.5~\mathrm{V}\leq V_{\mathrm{GS}}\leq0.5$~V at $0.1~\mathrm{V}\leq V_{\mathrm{DS}}\leq1.0$~V. Forward and reverse gate-voltage sweeps were measured under otherwise identical bias conditions, while the gate-leakage current was recorded simultaneously with the drain current. Throughout the transistor analysis, $I_{\mathrm{D}}$ and $I_{\mathrm{G}}$ denote the drain and gate currents normalized by the corresponding channel width, respectively.

The local subthreshold swing was calculated from
\[
SS
=
\left(
\frac{\mathrm{d}\log_{10}I_{\mathrm{D}}}
{\mathrm{d}V_{\mathrm{GS}}}
\right)^{-1},
\]
using the forward gate-voltage sweeps for all three gate-stack configurations. The minimum $SS$ within the experimentally resolved subthreshold regime is reported unless otherwise stated. The $SS$ characteristics in Fig.~\ref{Fig:FET}e and the device-to-device values in Extended Data Fig.~\ref{Fig:FET_Controls}h were evaluated at $V_{\mathrm{DS}}=0.1$~V. No smoothing was applied to the transfer characteristics before $SS$ extraction.

For the hysteresis analysis, the linear-extrapolation threshold voltage, $V_{\mathrm{th,LE}}$, was defined as the gate-voltage-axis intercept of the tangent to the $I_{\mathrm{D}}$--$V_{\mathrm{GS}}$ characteristic at the point of maximum transconductance on a linear current scale. With $g_{\mathrm{m}}=\mathrm{d}I_{\mathrm{D}}/\mathrm{d}V_{\mathrm{GS}}$ and $V_{\mathrm{GS}}^{*}$ denoting the gate voltage at this point,
\[
V_{\mathrm{th,LE}}
=
V_{\mathrm{GS}}^{*}
-
\frac{
I_{\mathrm{D}}(V_{\mathrm{GS}}^{*})
}{
g_{\mathrm{m}}(V_{\mathrm{GS}}^{*})
}.
\]
The hysteresis window was defined as the difference in $V_{\mathrm{th,LE}}$ between the forward and reverse sweeps at the same drain bias. The gate-voltage step size was 0.01~V and the effective sweep rate was 7.7~mV~s$^{-1}$. The same definition was used for the TIHGS-, BiF\textsubscript{3}-only-, and BiO\textsubscript{x}/Bi\textsubscript{2}Se\textsubscript{3}-gated devices.

Drain-induced barrier lowering ($DIBL$) was extracted from the forward subthreshold transfer characteristics using a constant-current method. We denote by $V_{\mathrm{GS,CC}}(V_{\mathrm{DS}};I_{\mathrm{ref}})$ the gate voltage at which the width-normalized drain current reaches the selected reference value, $I_{\mathrm{D}}=I_{\mathrm{ref}}$. These gate voltages were obtained by linear interpolation between adjacent measured points in $\log_{10}I_{\mathrm{D}}$ versus $V_{\mathrm{GS}}$. The signed coefficient was calculated as
\[
DIBL(I_{\mathrm{ref}})
=
\frac{
V_{\mathrm{GS,CC}}(0.1~\mathrm{V};I_{\mathrm{ref}})
-
V_{\mathrm{GS,CC}}(1.0~\mathrm{V};I_{\mathrm{ref}})
}{
1.0~\mathrm{V}-0.1~\mathrm{V}
}.
\]
The TIHGS- and BiF\textsubscript{3}-only-gated devices in Fig.~\ref{Fig:FET}f were evaluated at the common reference current $I_{\mathrm{ref}}=10$~pA~$\mu$m$^{-1}$ ($10^{-5}$~$\mu$A~$\mu$m$^{-1}$). The BiO\textsubscript{x}/Bi\textsubscript{2}Se\textsubscript{3} control in Extended Data Fig.~\ref{Fig:FET_Controls}g was evaluated at $I_{\mathrm{ref}}=100$~pA~$\mu$m$^{-1}$ ($10^{-4}$~$\mu$A~$\mu$m$^{-1}$), providing greater separation from its leakage-dominated low-current region.

The sign was retained to distinguish the direction of the drain-bias-induced shift: negative $DIBL$ indicates that a more positive gate voltage is required to maintain the selected current as the drain bias increases. Values in V~V$^{-1}$ were multiplied by $10^{3}$ for reporting in mV~V$^{-1}$. The coefficient quantifies the net subthreshold drain-bias dependence at the specified current criterion, while its magnitude describes the sensitivity irrespective of the shift direction.

For the device-to-device comparison in Extended Data Fig.~\ref{Fig:FET_Controls}h, the minimum $SS$ of each device was extracted from the forward transfer characteristic at $V_{\mathrm{DS}}=0.1$~V using the same local-derivative procedure. The gate-leakage metric was $|I_{\mathrm{G}}|$, evaluated at $V_{\mathrm{GS}}=0.5$~V and $V_{\mathrm{DS}}=0.1$~V for every configuration. Each plotted marker pairs the minimum $SS$ and gate-leakage metric of one independent device. Sample sizes are specified in Statistics and Reproducibility.

\subsection{Statistics and Reproducibility}
\label{M10}

No inferential statistical hypothesis tests were applied. For the MIM-capacitor statistics in Fig.~\ref{Fig:Buried_Boundary}f,g, the sample sizes were $n=25$, 33, 24, and 5 independent capacitors for BiF\textsubscript{3}-only (p20), TIHGS-13 (p20), TIHGS-43 (p50), and TIHGS-43 (c$N$), respectively. The Al\textsubscript{2}O\textsubscript{3} reference in Extended Data Fig.~\ref{Fig:MIM_Ref}a,b comprised $n=75$ independent capacitors. For the device-to-device $SS$--gate-leakage comparison in Extended Data Fig.~\ref{Fig:FET_Controls}h, the sample sizes were $n=4$ for the TIHGS-gated devices and $n=3$ each for the BiF\textsubscript{3}-only- and BiO\textsubscript{x}/Bi\textsubscript{2}Se\textsubscript{3}-gated devices. The $DIBL$ values in Fig.~\ref{Fig:FET}f and Extended Data Fig.~\ref{Fig:FET_Controls}g describe the representative devices shown, rather than averages over these device populations. Mean and median values are reported where indicated.

\subsection{Use of Large Language Models}
\label{M11}

The authors wrote the original manuscript draft. ChatGPT (OpenAI) was subsequently used for grammar correction and language refinement of author-written text, together with editorial feedback on manuscript organization and the presentation of the authors' results. ChatGPT was not used to generate the original manuscript draft. The authors reviewed all proposed edits, revised the text as needed, and approved the final wording, while independently evaluating the scientific interpretations and conclusions. The authors prepared all figures, including data plots and schematics. No large language models or generative AI tools were used to generate, plot, draw, modify, or enhance the visual content of these figures. The authors take full responsibility for the accuracy, integrity, and final content of the manuscript.

\section*{Data Availability}

The data supporting the findings of this study are available from the corresponding authors upon reasonable request.

\section*{Code Availability}

No custom code central to the conclusions of this study was developed.

\section*{Acknowledgements}
We acknowledge the Gwangju Institute of Science and Technology (GIST) Nanoinfra for Compound Semiconductors (G-NICS), the GIST Advanced Institute of Instrumental Analysis (GAIA), the Otaniemi Research Infrastructure (OtaNano Micronova Centre and OtaNano Nanomicroscopy Centre), and the Korea Photonics Technology Institute (KOPTI) for providing their cleanroom facilities and technical support.

\section*{Funding}
This work was supported by the National Research Foundation of Korea (NRF) Grant, funded by the Korean Government's Ministry of Science and ICT (MSIT) with Grant Nos.~RS-2026-25587884 (to H.H.Y.), NRF-2023R1A2C1006519 (to K.P.), RS-2026-25593368 (to S.B.C.) and RS-2025-25436243 (to M.S.); the Research Council of Finland with Grant Nos.~360411, 359009, 365686, 367808, and 374168 (to Z.S.); the Research Council of Finland Flagship Programme (Photonics Research and Innovation, PREIN) with Grant No.~320167 (to Z.S.); the Research Council of Finland Flagship Programme (Finnish Quantum Flagship, FQF) with Grant No.~358877; the Research Council of Finland (Finnish Centre of Excellence in Quantum Materials, QMAT) with Grant No.~86572 (to Z.S.); the Jane and Aatos Erkko Foundation and the Technology Industries of Finland Centennial Foundation with Grant No.~Future Makers 2022 (to Z.S.); the Global Value-Up 10X Project with Grant No.~KH1100, funded by the GIST in 2026 (to H.H.Y. and H.-J.S.); the GIST Graduate International Research Experience Fellowship (to H.H.Y. and G.J.); the U.S. Department of Energy, Office of Science, Basic Energy Sciences with Grant No.~DOE-SC0020653 (to S.A.T.); the U.S. National Science Foundation with Grant Nos.~CBET-2330110 (to S.A.T.), DMR-2534832 (to S.A.T.), and CMMI-2533623 (to S.A.T.); the Army Research Office with Grant No.~W911NF-25-1-0045 (to S.A.T.); the Applied Materials Inc. (to S.A.T.); and the Lawrence Semiconductor Research Laboratory Inc. (to S.A.T.).

\section*{Author Information}

\bmhead{Contributions}
M.S., J.K., M.G.U., and W.K. contributed equally to this work. H.H.Y. conceived the ideas and designed the experiments. S.H.C., D.-H.K., H.-J.S., C.-H.L., M.C., S.B.C., Z.S., K.P., and H.H.Y. proposed theoretical interpretations. H.U.L., D.W.J., and S.B.C. performed DFT calculations and SRIM simulations. S.A.T. provided high-quality Bi\textsubscript{2}Se\textsubscript{3}, MoS\textsubscript{2}, WSe\textsubscript{2}, and h-BN crystals. D.L. and W.S. offered wafer-scale Bi\textsubscript{2}Se\textsubscript{3} films. M.S., M.G.U., W.K., J.P., X.C., and H.H.Y. developed and optimized the SF\textsubscript{6}/Ar plasma process for controlled surface conversion of Bi\textsubscript{2}Se\textsubscript{3} into BiF\textsubscript{3}. M.S., M.G.U., W.K., J.P., D.L., G.J., S.M., and H.H.Y. fabricated the devices. L.Y. obtained TEM images. J.L. carried out XPS. M.S., W.K., J.P., S.C., K.S.K., Z.S., and H.H.Y. characterized the devices using TEM, HAADF-STEM, EDS, AFM, Raman, and XPS datasets. M.S., J.K., W.K., J.P., K.P., and H.H.Y. performed the electrical measurements. G.J., S.M., Y.T.M., and X.C. helped with the electrical measurements. Y.D., S.H.C., K.S.K., D.-H.K., H.-J.S., W.S., S.A.T., C.-H.L., M.C., S.B.C., Z.S., and K.P. commented on the experimental results and assisted with the data analysis. Y.D., D.-H.K., H.-J.S., M.C., S.B.C., Z.S., and K.P. suggested insights for organizing the data. H.H.Y. wrote the manuscript. S.B.C., Z.S., K.P., and H.H.Y. supervised the research. All authors participated extensively in the scientific discussion and contributed to the manuscript.

\bmhead{Corresponding authors}
Correspondence to Sung Beom Cho, Zhipei Sun, Kibog Park, and Hoon Hahn Yoon

\section*{Ethics Declarations}

\bmhead{Competing interests}
The authors declare no competing interests.


\clearpage

\section*{Extended Data Figures and Tables}

\setcounter{figure}{0}
\renewcommand{\thefigure}{\arabic{figure}}
\renewcommand{\figurename}{Extended Data Fig.}

\ifdefined\theHfigure
  \renewcommand{\theHfigure}{ED.\arabic{figure}}
\fi

\newcounter{extendeddatatable}
\renewcommand{\theextendeddatatable}{\arabic{extendeddatatable}}


\begin{figure}[H]
\centering
\includegraphics[width=\textwidth]{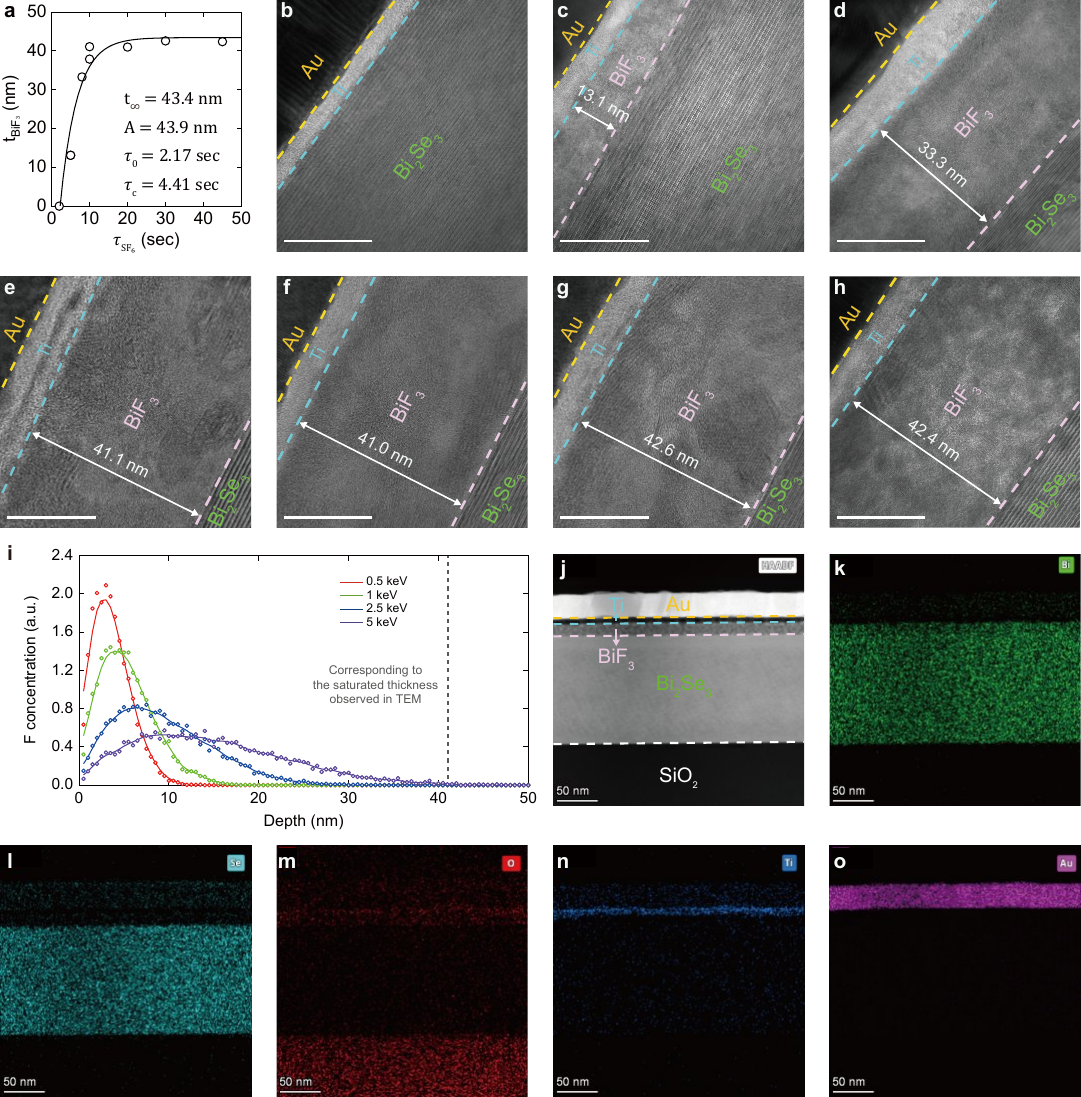}
\caption{\textbf{Self-limited surface conversion and fluorine-penetration length scale.}
\textbf{a}, Converted BiF\textsubscript{3} thickness ($t_{\mathrm{BiF_3}}$) as a function of SF\textsubscript{6} plasma-treatment time ($\tau_{\mathrm{SF_6}}$). The line is an empirical saturating-exponential fit used to parameterize the observed approach to a limiting thickness; the fitted time constant is not assigned a unique microscopic kinetic meaning.
\textbf{b--h}, Cross-sectional transmission electron microscopy (TEM) images after plasma exposure for 2, 5, 8, 10, 20, 30 and 45~s, respectively. Scale bars, 20~nm.
\textbf{i}, Fluorine depth distributions in Bi\textsubscript{2}Se\textsubscript{3} calculated using Stopping and Range of Ions in Matter (SRIM) simulations for assumed F\textsuperscript{+} incident energies of 0.5, 1, 2.5 and 5~keV. The dashed line marks the experimentally observed $\sim$40--43-nm saturation scale. The 5-keV case is included as a deliberately high-energy upper-bound calculation; the simulations constrain an accessible ballistic penetration length scale and do not represent the measured plasma ion-energy distribution or fluorination kinetics.
\textbf{j--o}, High-angle annular dark-field scanning transmission electron microscopy (HAADF-STEM) image \textbf{(j)} and corresponding energy-dispersive X-ray spectroscopy (EDS) elemental maps of Bi \textbf{(k)}, Se \textbf{(l)}, O \textbf{(m)}, Ti \textbf{(n)} and Au \textbf{(o)}. Se is concentrated predominantly in the residual Bi\textsubscript{2}Se\textsubscript{3}, whereas the converted region remains Bi-containing. Scale bars, 50~nm.}
\label{Fig:TEM_SRIM_EDS}
\end{figure}


\begin{figure}[H]
\centering
\includegraphics[width=\textwidth]{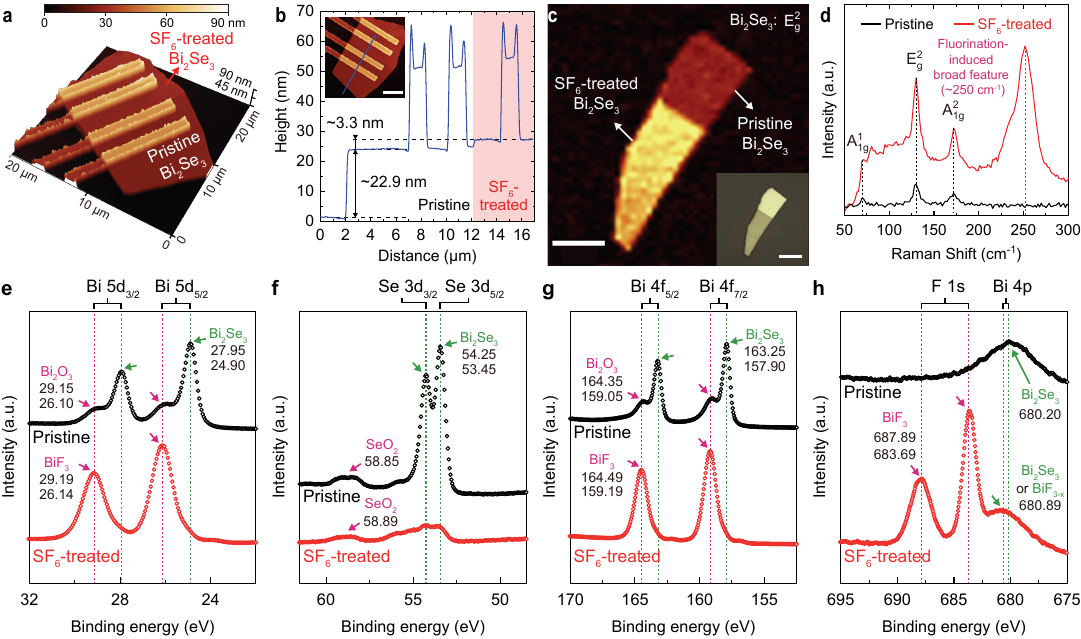}
\caption{\textbf{Spatially selective conversion of Bi\textsubscript{2}Se\textsubscript{3} into BiF\textsubscript{3}.}
\textbf{a,b}, Atomic force microscopy (AFM) topography across pristine and SF\textsubscript{6}-treated Bi\textsubscript{2}Se\textsubscript{3} regions \textbf{(a)} and corresponding line-height profile \textbf{(b)}, showing a surface-height increase of $\sim$3.3~nm after fluorination. The difference between the converted-layer thickness ($\sim$13.1~nm), independently determined by cross-sectional TEM under the same 5-s treatment condition, and the measured surface-height increase ($\sim$3.3~nm) indicates that the converted layer cannot be explained by additive deposition and is consistent with consumption and volumetric expansion of the underlying Bi\textsubscript{2}Se\textsubscript{3}. Scale bar of inset in \textbf{b}, 5~$\mu$m.
\textbf{c}, Raman map of the Bi\textsubscript{2}Se\textsubscript{3} $\mathrm{E}^{2}_{g}$ mode across pristine and treated regions; inset, corresponding optical micrograph. Scale bar, 7~$\mu$m; inset scale bar, 10~$\mu$m.
\textbf{d}, Raman spectra showing the characteristic Bi\textsubscript{2}Se\textsubscript{3} modes and an additional broad feature near 250~cm$^{-1}$ after fluorination.
\textbf{e--h}, X-ray photoelectron spectroscopy (XPS) of pristine and treated Bi\textsubscript{2}Se\textsubscript{3}: Bi 5d \textbf{(e)}, Se 3d \textbf{(f)}, Bi 4f \textbf{(g)}, and F 1s/Bi 4p \textbf{(h)}. Fluorination suppresses the Bi\textsubscript{2}Se\textsubscript{3}-derived Bi and Se components, introduces Bi--F components in the Bi spectra, and produces an F 1s feature near 687.9~eV, supporting formation of a fluorinated Bi-containing surface phase.}
\label{Fig:AFM_Raman_XPS}
\end{figure}


\begin{figure}[H]
\centering
\includegraphics[width=\textwidth]{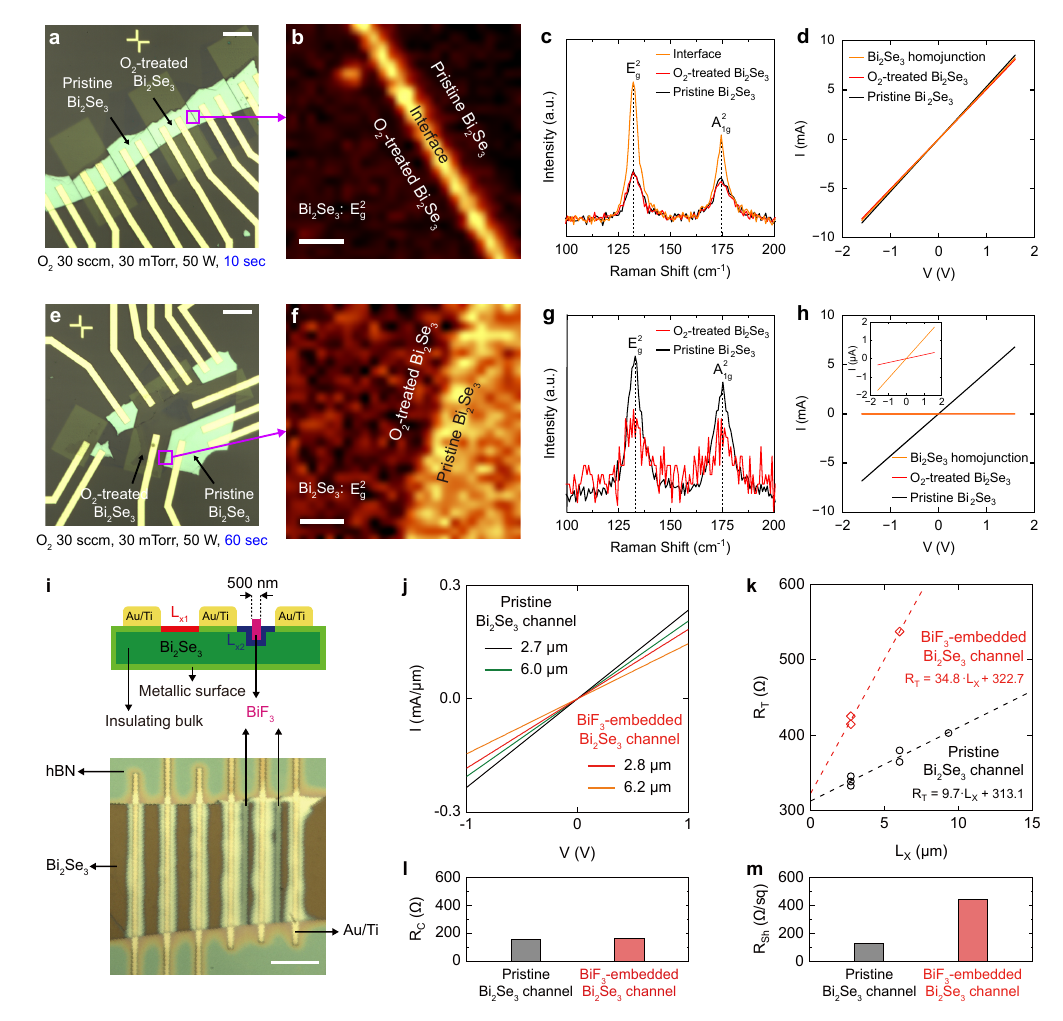}
\caption{\textbf{O\textsubscript{2}-plasma control and quantitative buried conduction in Bi\textsubscript{2}Se\textsubscript{3}.}
\textbf{a--d}, Bi\textsubscript{2}Se\textsubscript{3} after 10~s of O\textsubscript{2}-plasma treatment: optical micrograph \textbf{(a)}, Raman map of the $\mathrm{E}^{2}_{g}$ mode \textbf{(b)}, Raman spectra from pristine, treated, and interfacial regions \textbf{(c)}, and current--voltage ($I$--$V$) characteristics of pristine, treated, and lateral pristine/treated structures \textbf{(d)}. Scale bars, 50~$\mu$m in \textbf{a} and 1~$\mu$m in \textbf{b}.
\textbf{e--h}, Corresponding measurements after 60~s of O\textsubscript{2}-plasma treatment: optical micrograph \textbf{(e)}, Raman map \textbf{(f)}, Raman spectra \textbf{(g)}, and $I$--$V$ characteristics \textbf{(h)}. The treated region remains conductive, unlike after SF\textsubscript{6} fluorination. Scale bars, 50~$\mu$m in \textbf{e} and 1~$\mu$m in \textbf{f}. O\textsubscript{2}-plasma treatment was performed at 30~sccm, 30~mTorr, and 50~W.
\textbf{i--m}, Transfer-length-method (TLM) analysis of conduction through residual Bi\textsubscript{2}Se\textsubscript{3} beneath a locally converted BiF\textsubscript{3} region. \textbf{i}, Schematic and optical micrograph of the TLM geometry containing a $\sim$500-nm-wide converted region; scale bar, 5~$\mu$m. \textbf{j}, Representative $I$--$V$ characteristics. \textbf{k}, Total resistance versus channel length, with linear fits $R_{\mathrm{T}}=9.7L_{\mathrm{X}}+313.1$ and $R_{\mathrm{T}}=34.8L_{\mathrm{X}}+322.7$ for pristine and converted structures, respectively. \textbf{l,m}, Extracted contact resistance and sheet resistance. The contact resistance remains $\sim$157--161~$\Omega$, whereas the sheet resistance increases by $\sim$3.5-fold after local conversion.}
\label{Fig:Electrical_Controls}
\end{figure}


\begin{figure}[H]
\centering
\includegraphics[width=\textwidth]{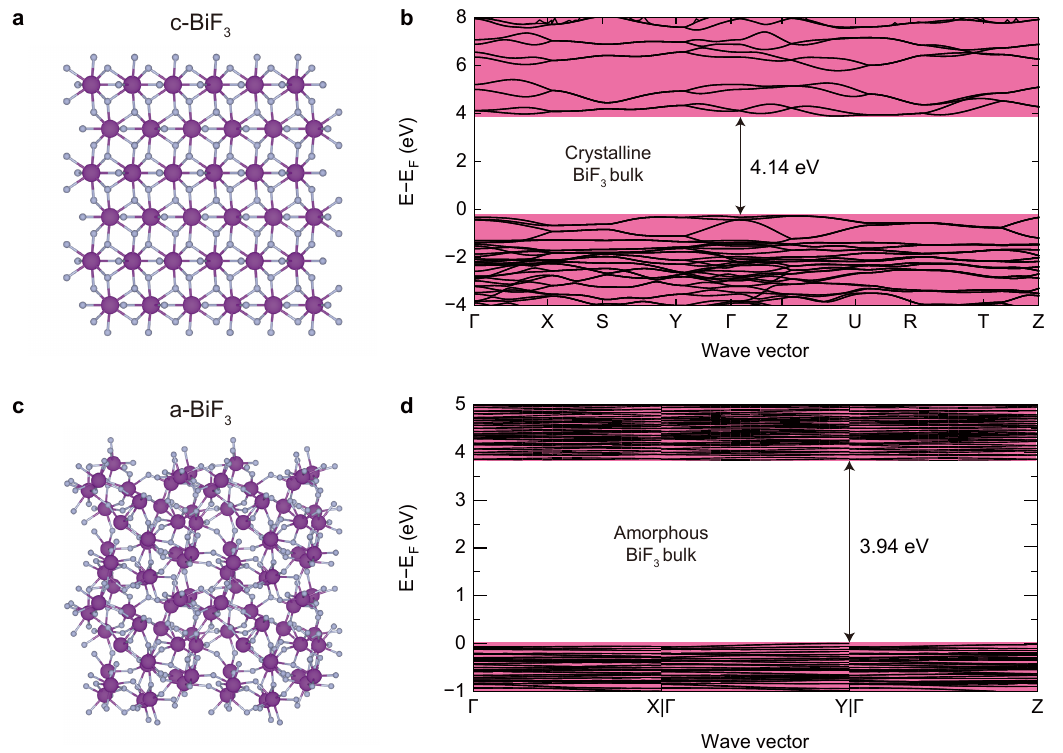}
\caption{\textbf{Calculated electronic band structures of crystalline and amorphous BiF\textsubscript{3}.}
\textbf{a}, Structural model of crystalline BiF\textsubscript{3}.
\textbf{b}, Calculated electronic band structure of crystalline BiF\textsubscript{3} along the $\Gamma$--X--S--Y--$\Gamma$--Z--U--R--T--Z high-symmetry path, showing an insulating band gap of 4.14~eV.
\textbf{c}, Structural model of amorphous BiF\textsubscript{3}.
\textbf{d}, Calculated electronic band structure of amorphous BiF\textsubscript{3} along the $\Gamma$--X, $\Gamma$--Y, and $\Gamma$--Z reciprocal-space directions, showing an insulating band gap of 3.94~eV. Energies in \textbf{b} and \textbf{d} are plotted relative to the calculated Fermi level ($E_{\mathrm{F}}$).}
\label{Fig:BiF3_DFT}
\end{figure}


\begin{table*}[!b]
\centering

\refstepcounter{extendeddatatable}
\label{tab:bandgap_permittivity}

\noindent\parbox{\linewidth}{%
\textbf{Extended Data Table~\theextendeddatatable\quad}
\textbf{Calculated band gaps and dielectric properties of Bi$_2$Se$_3$, BiF$_3$, and Bi$_2$O$_3$ in crystalline and amorphous phases.}
The static relative permittivity [$\varepsilon_{\mathrm{r}}(0)$] includes both electronic and ionic contributions, whereas the high-frequency relative permittivity [$\varepsilon_{\mathrm{r}}(\infty)$] represents the ion-clamped electronic contribution. The refractive index ($n$) was calculated from the averaged $\varepsilon_{\mathrm{r}}(\infty)$. Stoichiometric Bi$_2$O$_3$ is included only as a theoretical reference for the oxidized BiO$_x$ phase in the BiO$_x$/Bi$_2$Se$_3$ control gate stack demonstrated in Extended Data Fig.~\ref{Fig:FET_Controls}.
}

\vspace{0.8em}

\small
\renewcommand{\arraystretch}{1.1}

\makebox[\linewidth][c]{%
\begin{tabular}{cccccc}
\toprule

\multirow{2}{*}{Phase} &
\multirow{2}{*}{Material} &
\multirow{2}{*}{\begin{tabular}{c}Band gap\\(eV)\end{tabular}} &
\multicolumn{2}{c}{Relative permittivity} &
\multirow{2}{*}{$n$} \\

\cmidrule(lr){4-5}

& &
&
$\varepsilon_{\mathrm{r}}(0)$ &
$\varepsilon_{\mathrm{r}}(\infty)$ &
\\

\midrule

Crystalline &
Bi$_2$Se$_3$ &
0.310 &
38.057 &
37.851 &
6.152 \\

Amorphous &
Bi$_2$Se$_3$ &
0.777 &
42.715 &
13.329 &
3.651 \\

\midrule

Crystalline &
BiF$_3$ &
4.140 &
29.385 &
3.227 &
1.796 \\

Amorphous &
BiF$_3$ &
3.936 &
25.311 &
4.056 &
2.014 \\

\midrule

Crystalline &
Bi$_2$O$_3$ &
2.038 &
32.023 &
6.110 &
2.472 \\

Amorphous &
Bi$_2$O$_3$ &
2.096 &
34.938 &
6.118 &
2.473 \\

\bottomrule
\end{tabular}
}

\end{table*}


\begin{figure}[H]
\centering
\includegraphics[width=\textwidth]{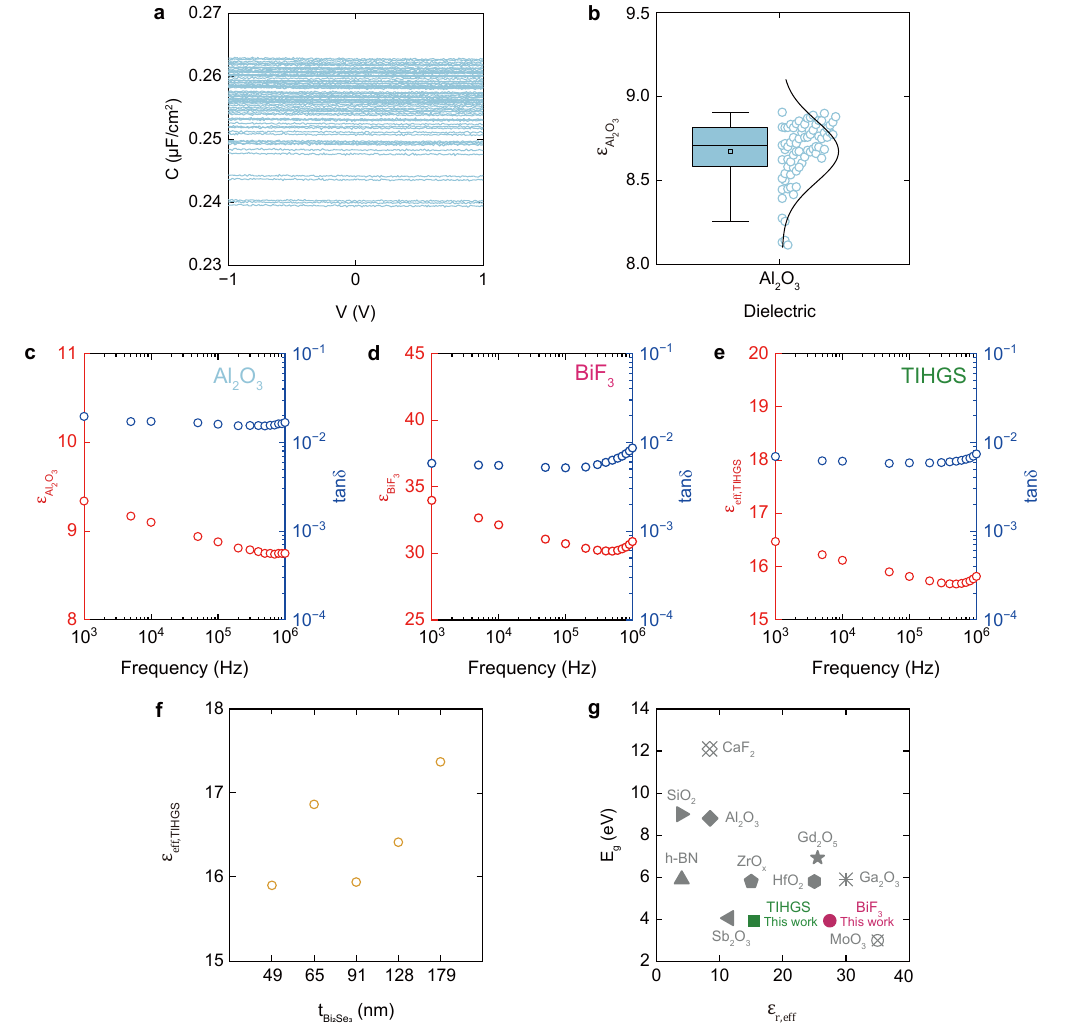}
\caption{\textbf{Reference dielectric response, thickness dependence and dielectric-property context of TIHGS electrostatics.}
\textbf{a}, Capacitance--voltage ($C$--$V$) characteristics of Au/Ag/Pt/Al\textsubscript{2}O\textsubscript{3}/Au/Ti MIM reference capacitors measured at 1~MHz with a 0.05-V AC excitation ($n=75$).
\textbf{b}, Distribution of the extracted Al\textsubscript{2}O\textsubscript{3} relative permittivity; mean and median values are 8.67 and 8.71. This reference response was used to de-embed the Al\textsubscript{2}O\textsubscript{3} series contribution from the polycrystalline capacitor structures in Fig.~\ref{Fig:Buried_Boundary}f,g.
\textbf{c--e}, Frequency-dependent relative permittivity and loss tangent ($\tan\delta$) of 30-nm Al\textsubscript{2}O\textsubscript{3} \textbf{(c)}, 30-nm Al\textsubscript{2}O\textsubscript{3}/23-nm BiF\textsubscript{3}-only (p20) \textbf{(d)}, and 30-nm Al\textsubscript{2}O\textsubscript{3}/TIHGS-43 (p50) \textbf{(e)}.
\textbf{f}, Effective relative permittivity $\varepsilon_{\mathrm{eff,TIHGS}}$ of TIHGS-43 (c$N$) versus nominal starting c-Bi\textsubscript{2}Se\textsubscript{3} thickness. $\varepsilon_{\mathrm{eff,TIHGS}}$ remains $\sim$15.9--17.4 as the starting thickness varies from $\sim$49 to $\sim$179~nm, without systematic thickness dependence.
\textbf{g}, Band gap ($E_g$) versus relative permittivity for BiF\textsubscript{3}, the TIHGS and representative gate dielectrics~\cite{jung2025advances}. For the TIHGS, the abscissa is the BiF\textsubscript{3}-thickness-normalized stack response $\varepsilon_{\mathrm{eff,TIHGS}}$ and the ordinate is the amorphous-BiF\textsubscript{3} band gap; the marker therefore provides dielectric-property context rather than an intrinsic material dielectric constant or $EOT$ advantage.}
\label{Fig:MIM_Ref}
\end{figure}


\begin{figure}[H]
\centering
\includegraphics[width=\textwidth]{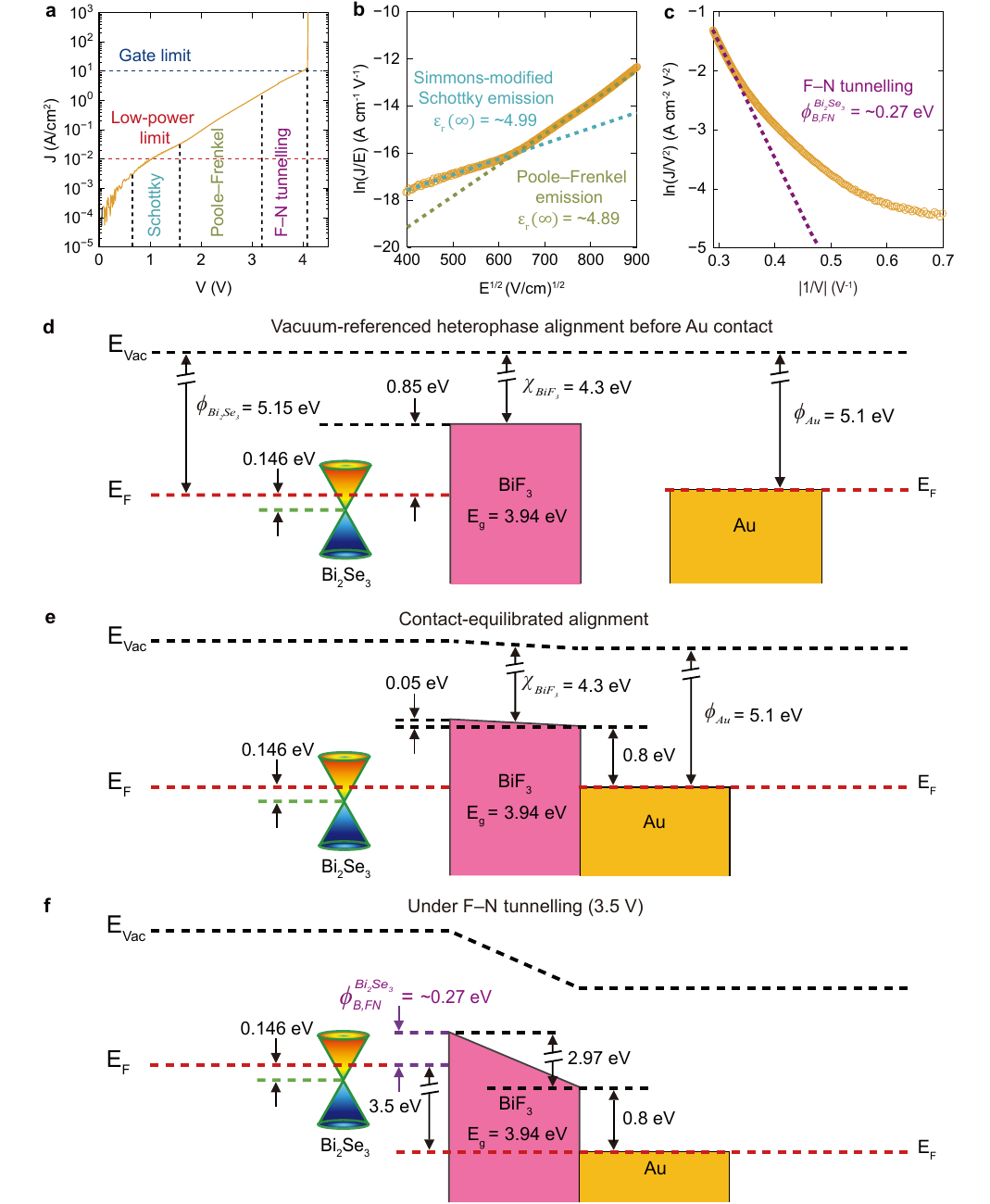} 
\caption{\textbf{Field-assisted leakage and band alignment of the TIHGS MIM capacitor.}
\textbf{a}, Current-density--voltage ($J$--$V$) characteristic of an Au/BiF\textsubscript{3}/Bi\textsubscript{2}Se\textsubscript{3}/Au/Ti TIHGS MIM capacitor, with the low-power operating range and higher-field transport regimes indicated.
\textbf{b}, Simmons-modified Schottky and Poole--Frenkel representations, yielding effective high-frequency relative permittivities of $\sim$4.99 and $\sim$4.89, respectively.
\textbf{c}, Fowler--Nordheim representation of the high-field regime; the linear region gives a model-dependent effective injection barrier $\Phi_{\mathrm{B,FN}}\approx0.27$~eV.
\textbf{d}, Vacuum-referenced heterophase alignment before Au contact, constructed from $\phi_{\mathrm{Bi_2Se_3}}=5.15$~eV and $\chi_{\mathrm{BiF_3}}=4.3$~eV, giving a nominal 0.85-eV separation. The reconstructed Bi\textsubscript{2}Se\textsubscript{3}-derived feature is shown $\sim$0.146~eV below $E_{\mathrm{F}}$. This reference construction does not independently resolve interface dipoles, chemical reconstruction or band bending.
\textbf{e}, Contact-equilibrated schematic after Au contact. With $\phi_{\mathrm{Au}}=5.1$~eV, the Au/Bi\textsubscript{2}Se\textsubscript{3} work-function difference is $\sim$0.05~eV and the vacuum-referenced Au/BiF\textsubscript{3} separation $\sim$0.8~eV; these values are not interpreted as independently measured equilibrium band offsets.
\textbf{f}, High-field band profile at $V=3.5$~V. Barrier tilting enables tunnelling towards BiF\textsubscript{3} conduction-band states, with the Fowler--Nordheim slope yielding an effective Bi\textsubscript{2}Se\textsubscript{3}-to-BiF\textsubscript{3} injection barrier of $\sim$0.27~eV.}
\label{Fig:TIHGS_Band}
\end{figure}


\begin{figure}[H]
\centering
\includegraphics[width=\textwidth]{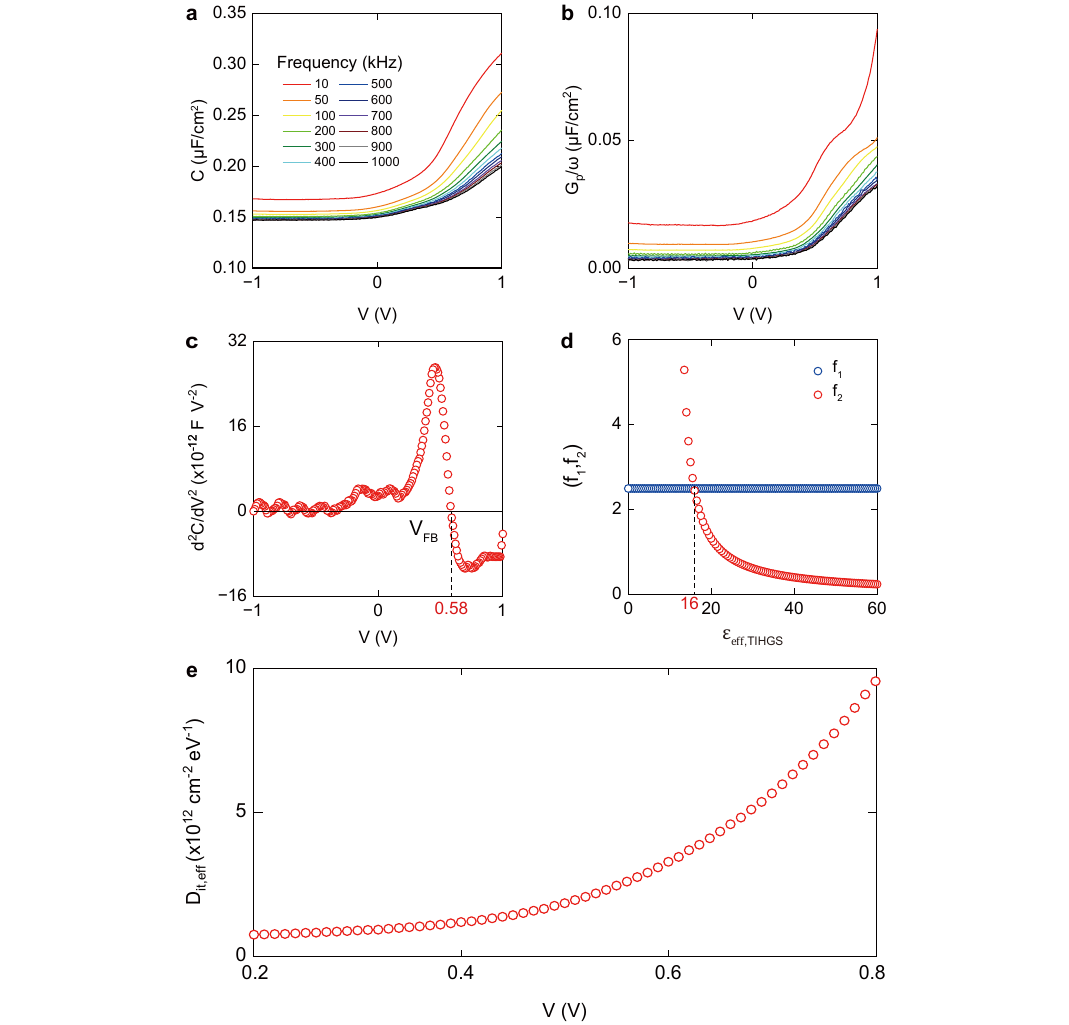}
\caption{\textbf{Semiconductor-integrated electrostatics of a WSe\textsubscript{2}/TIHGS MISCAP.}
\textbf{a}, Capacitance--voltage ($C$--$V$) characteristics of an Au/Ag/Pt/WSe\textsubscript{2}/BiF\textsubscript{3}/Bi\textsubscript{2}Se\textsubscript{3}/Au/Ti MISCAP measured from 10~kHz to 1~MHz over the gate-voltage range $-1~\mathrm{V}\leq V_{\mathrm{G}}\leq1~\mathrm{V}$ using an AC excitation amplitude of 0.05~V.
\textbf{b}, Corresponding parallel-conductance response, $G_{\mathrm{p}}/\omega$, as a function of gate voltage at different measurement frequencies. No well-resolved $G_{\mathrm{p}}/\omega$ maximum suitable for conventional conductance-method extraction is observed within the accessible low-leakage bias window.
\textbf{c}, Second derivative of the 10-kHz $C$--$V$ characteristic used to determine the flat-band voltage. The physically relevant zero crossing associated with the principal accumulation-to-depletion inflection gives $V_{\mathrm{FB}}\approx0.58$~V. At higher frequencies, the corresponding transition extends beyond the accessible low-leakage bias range and is therefore not used for flat-band extraction.
\textbf{d}, Graphical consistency analysis of the TIHGS response at the independently determined $V_{\mathrm{FB}}$. The experimentally determined function $f_1$ is compared with $f_2$ calculated as a function of trial $\varepsilon_{\mathrm{eff,TIHGS}}$, with the intersection yielding $\varepsilon_{\mathrm{eff,TIHGS}}\approx16$.
\textbf{e}, Effective electrically active trap density ($D_{\mathrm{it,eff}}$) obtained from the 10-kHz and 1-MHz $C$--$V$ characteristics using the high--low-frequency capacitance method. Quantitative interpretation is restricted to the bias-dependent depletion-to-accumulation transition, excluding the nearly gate-voltage-independent depletion floor and strong-accumulation regimes. The minimum value within the selected analysis range is $\sim7.48\times10^{11}$~cm$^{-2}$~eV$^{-1}$.}
\label{Fig:MISCAP}
\end{figure}

\clearpage


\begingroup
\centering

\refstepcounter{extendeddatatable}
\label{tab:dielectric_stack}

\noindent\parbox{\linewidth}{%
\textbf{Extended Data Table~\theextendeddatatable\quad}
\textbf{Electrostatic response of BiF$_3$ and TIHGS structures and parameters extracted from the WSe$_2$/TIHGS MISCAP.}
\textbf{a}, Experimentally extracted dielectric response of BiF$_3$ and TIHGS capacitor structures. For TIHGS structures, $\varepsilon_{\mathrm{eff,TIHGS}}$ denotes the BiF$_3$-thickness-normalized effective relative permittivity of the complete BiF$_3$/Bi$_2$Se$_3$ electrostatic boundary system and should not be interpreted as the intrinsic dielectric constant of BiF$_3$. For the fully converted BiF$_3$-only (p20) reference, $\varepsilon_{\mathrm{BiF_3}}$ was extracted using the independently measured post-conversion BiF$_3$ thickness of $\sim$23~nm. TIHGS-13 and TIHGS-43 denote structures containing $\sim$13- and $\sim$43-nm-thick converted BiF$_3$ layers, respectively; p20 and p50 denote nominal starting polycrystalline Bi$_2$Se$_3$ thicknesses of 20 and 50~nm, respectively; and c$N$ denotes crystalline Bi$_2$Se$_3$ with nominal starting thickness $N$.
\textbf{b}, Electrostatic parameters extracted from the Au/Ag/Pt/WSe$_2$/BiF$_3$/Bi$_2$Se$_3$/Au/Ti MISCAP. Flat-band quantities were obtained from the 10-kHz $C$--$V$ characteristic. $C_{\mathrm{g}}$ denotes the effective TIHGS capacitance corresponding to $\varepsilon_{\mathrm{eff,TIHGS}}\approx16$ and a $\sim$43-nm-thick converted BiF$_3$ layer. $L_{\mathrm{D,eff}}$ and $N_{\mathrm{app}}$ denote the effective semiconductor screening length and apparent majority-carrier concentration, respectively, evaluated using $\varepsilon_{\perp,\mathrm{WSe_2}}=4.2$, as adopted in continuum electrostatic modelling of bulk-like WSe\textsubscript{2} flakes~\cite{yu2017photogenerated}. $D_{\mathrm{it,eff}}^{\mathrm{min}}$ denotes the minimum effective electrically active trap density extracted from the 10-kHz/1-MHz high--low-frequency analysis within the selected bias-dependent transition region. Detailed extraction procedures and limitations are provided in Supplementary Notes~\ref{SN:Flat-Band_Screening} and \ref{SN:Interface-Trap}.
}

\vspace{0.9em}

\noindent\textbf{a, Dielectric response of BiF$_3$ and TIHGS}

\vspace{0.4em}

\small
\renewcommand{\arraystretch}{1.30}
\setlength{\tabcolsep}{4pt}

\makebox[\linewidth][c]{%
\begin{tabular}{
@{}
C{0.27\textwidth}
C{0.35\textwidth}
C{0.20\textwidth}
C{0.12\textwidth}
@{}
}

\toprule

\textbf{Structure} &
\textbf{Sample configuration} &
\textbf{Electrostatic quantity} &
\textbf{Value} \\

\midrule

\multirow{2}{*}{
    \makecell[c]{
        Crystalline\\
        TIHGS MIM
    }
}
&
\multirow{2}{*}{
    \makecell[c]{
        TIHGS-43 (c$N$)\\
        $N=49$, 65, 91, 128, or 179~nm
    }
}
&
Mean $\varepsilon_{\mathrm{eff,TIHGS}}$
&
\textbf{16.50}
\\

&
&
Median $\varepsilon_{\mathrm{eff,TIHGS}}$
&
\textbf{16.41}
\\[0.8ex]

\makecell[c]{
    WSe$_2$/TIHGS\\
    MISCAP
}
&
\makecell[c]{
    Au/Ag/Pt/WSe$_2$/a-BiF$_3$/\\
    c-Bi$_2$Se$_3$/Au/Ti\\
    ($\sim$43-nm converted BiF$_3$)
}
&
$\varepsilon_{\mathrm{eff,TIHGS}}$
&
\textbf{$\sim$16}
\\[0.8ex]

\multirow{4}{*}{
    \makecell[c]{
        Polycrystalline\\
        TIHGS MIM
    }
}
&
\multirow{2}{*}{
    \makecell[c]{
        TIHGS-13 (p20)
    }
}
&
Mean $\varepsilon_{\mathrm{eff,TIHGS}}$
&
\textbf{16.12}
\\

&
&
Median $\varepsilon_{\mathrm{eff,TIHGS}}$
&
\textbf{16.10}
\\[0.8ex]

&
\multirow{2}{*}{
    \makecell[c]{
        TIHGS-43 (p50)
    }
}
&
Mean $\varepsilon_{\mathrm{eff,TIHGS}}$
&
\textbf{14.83}
\\

&
&
Median $\varepsilon_{\mathrm{eff,TIHGS}}$
&
\textbf{14.72}
\\[0.8ex]

\multirow{2}{*}{
    \makecell[c]{
        Fully converted\\
        BiF$_3$-only MIM
    }
}
&
\multirow{2}{*}{
    \makecell[c]{
        BiF$_3$-only (p20)\\
        ($t_{\mathrm{BiF_3}}\approx23$~nm)
    }
}
&
Mean $\varepsilon_{\mathrm{BiF_3}}$
&
\textbf{28.65}
\\

&
&
Median $\varepsilon_{\mathrm{BiF_3}}$
&
\textbf{30.16}
\\

\bottomrule
\end{tabular}
}

\vspace{1.2em}

\noindent\textbf{b, Electrostatic parameters extracted from the WSe$_2$/TIHGS MISCAP}

\vspace{0.4em}

\renewcommand{\arraystretch}{1.25}
\setlength{\tabcolsep}{5pt}

\makebox[\linewidth][c]{%
\begin{tabular}{ccccccc}

\toprule

$t_{\mathrm{WSe_2}}$ (nm) &
$V_{\mathrm{FB}}$ (V) &
\makecell{$C_{\mathrm{FB}}$\\($10^{-7}$ F cm$^{-2}$)} &
\makecell{$C_{\mathrm{g}}$\\($10^{-7}$ F cm$^{-2}$)} &
$L_{\mathrm{D,eff}}$ (nm) &
\makecell{$N_{\mathrm{app}}$\\($10^{17}$ cm$^{-3}$)} &
\makecell{$D_{\mathrm{it,eff}}^{\mathrm{min}}$\\($10^{11}$ cm$^{-2}$ eV$^{-1}$)} \\

\midrule

37.37 &
0.58 &
2.34 &
3.29 &
4.60 &
2.83 &
7.48 \\

\bottomrule

\end{tabular}
}

\endgroup

\clearpage


\begin{figure}[H]
\centering
\includegraphics[width=\textwidth]{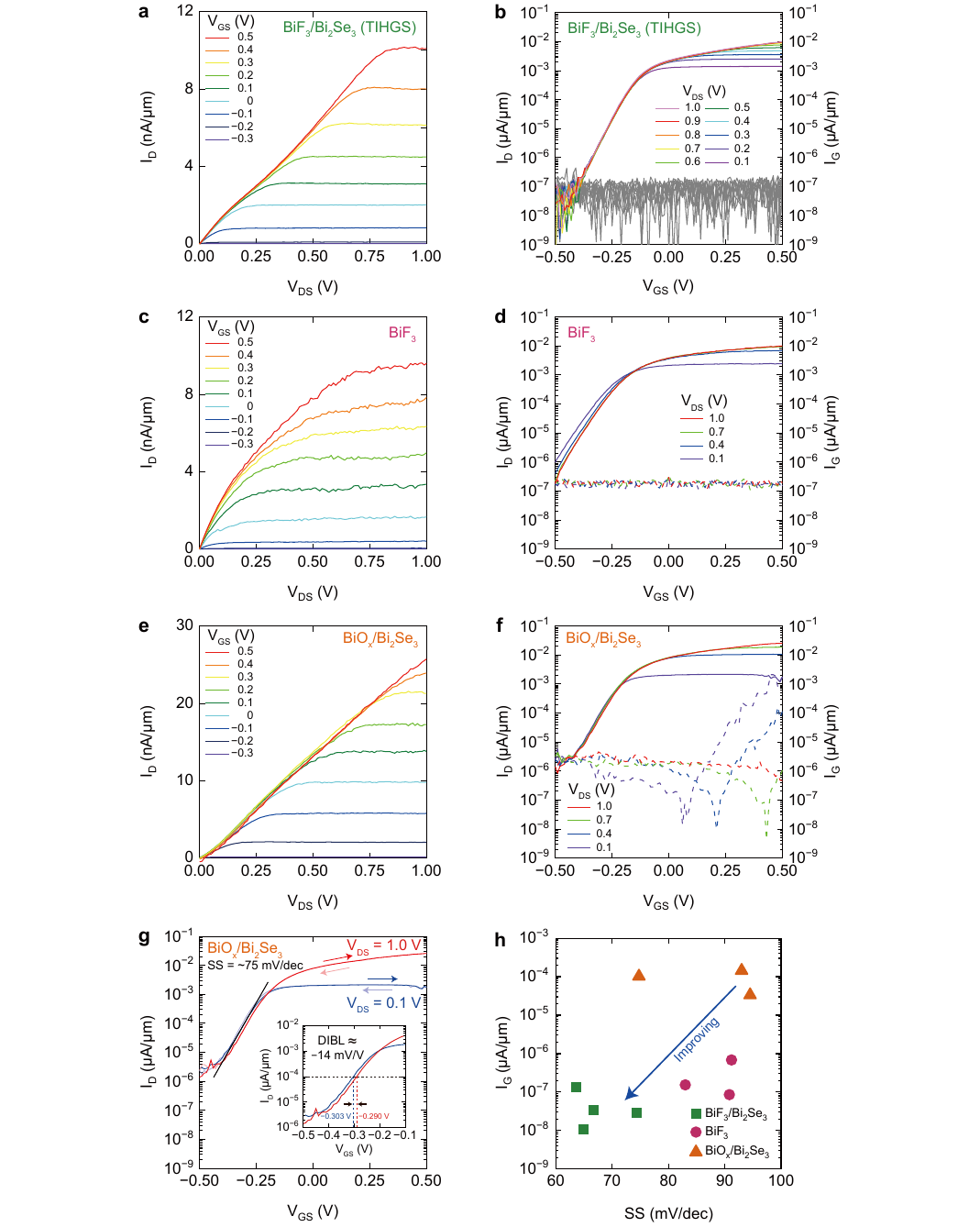}
\caption{\textbf{Extended transistor characteristics, control devices and device-to-device comparison.}
\textbf{a--f}, Output characteristics \textbf{(a,c,e)} and drain-bias-dependent transfer characteristics \textbf{(b,d,f)} of TIHGS-gated \textbf{(a,b)}, BiF\textsubscript{3}-only-gated \textbf{(c,d)} and air-oxidized BiO\textsubscript{x}/Bi\textsubscript{2}Se\textsubscript{3}-gated \textbf{(e,f)} MoS\textsubscript{2} FETs. Simultaneously measured gate-leakage currents are shown in \textbf{b,d,f}.
\textbf{g}, Forward and reverse transfer characteristics of the BiO\textsubscript{x}/Bi\textsubscript{2}Se\textsubscript{3}-gated control at $V_{\mathrm{DS}}=0.1$ and 1.0~V, complementing the corresponding TIHGS and BiF\textsubscript{3}-only measurements in Fig.~\ref{Fig:FET}b,d. Arrows indicate the gate-voltage sweep direction. The device exhibits a minimum $SS$ of $\sim75$~mV~dec$^{-1}$ and an on--off current ratio of $\sim10^{4}$. Inset, constant-current extraction from the forward sweeps at $I_{\mathrm{D}}=100$~pA~$\mu$m$^{-1}$, yielding a signed $DIBL\approx-14$~mV~V$^{-1}$.
\textbf{h}, Device-to-device comparison of minimum $SS$ extracted from forward transfer characteristics at $V_{\mathrm{DS}}=0.1$~V and width-normalized gate leakage evaluated at $V_{\mathrm{GS}}=0.5$~V and $V_{\mathrm{DS}}=0.1$~V.}
\label{Fig:FET_Controls}
\end{figure}

\clearpage


\section*{Supplementary Information}

\newcounter{suppnote}
\renewcommand{\thesuppnote}{\arabic{suppnote}}

\refstepcounter{suppnote}
\subsection*{Supplementary Note~\thesuppnote. Screening-Boundary Relocation as an Electrostatic Design Variable}
\label{SN:Screening_Boundary}

A gate stack is commonly described by the capacitance of its bulk dielectric, but the gate electric field is ultimately screened at finite electronic and interfacial boundaries. For the two-dimensional (2D) semiconductor transistors considered here, a conventional stack can be represented as
\[
\frac{1}{C_{\mathrm{g,conv}}}
=
\frac{1}{C_{\mathrm{d,g}}}
+
\frac{1}{C_{\mathrm{ins}}}
+
\frac{1}{C_{\mathrm{d,ch}}}
+
\frac{1}{C_{\mathrm{vdW}}},
\]
where $C_{\mathrm{d,g}}$ denotes the effective gate-side interfacial response of the electrode/dielectric boundary, $C_{\mathrm{d,ch}}$ denotes the corresponding dielectric/channel interfacial response, and $C_{\mathrm{vdW}}$ denotes the geometric van der Waals (vdW)-gap contribution. In particular, $C_{\mathrm{d,g}}$ can incorporate finite electrode screening together with interface-specific bonding, polarization, chemical reconstruction and structural disorder. These gate-side contributions need not scale proportionally with the physical dielectric thickness and can therefore become increasingly important as the bulk-insulator capacitance increases during equivalent oxide thickness ($EOT$) scaling~\cite{lau2023dielectrics,cao2023future,jung2025advances,kim2025gate,pourfath2026device}. The bias-dependent semiconductor-body response is treated separately from the gate-stack contribution defined above. Thus, even for an intrinsically high-$\kappa$ dielectric, the complete gate response depends on the electronic character and spatial position of the boundaries at which screening charge is established.

The topological-insulator heterophase gate stack (TIHGS) changes this gate-side boundary condition by separating dielectric isolation from the location of electronic screening. Amorphous BiF\textsubscript{3} (a-BiF\textsubscript{3}) provides the insulating barrier facing the semiconductor channel, whereas residual conducting crystalline Bi\textsubscript{2}Se\textsubscript{3} (c-Bi\textsubscript{2}Se\textsubscript{3}) remains electrically coupled to the gate electrode. Position-resolved calculations further show that the structural a-BiF\textsubscript{3}/c-Bi\textsubscript{2}Se\textsubscript{3} interface is gap-opened while an electronically active gap-closed state is reconstructed in the adjacent c-Bi\textsubscript{2}Se\textsubscript{3} subinterface layer. We therefore use the term ``buried screening boundary'' for this heterophase-associated electronic boundary region rather than for an ideal atomically sharp chemical interface. For the 2D devices studied here, the corresponding stack contribution is
\[
\frac{1}{C_{\mathrm{g,TIHGS}}}
=
\frac{1}{C_Q}
+
\frac{1}{C_{\mathrm{BiF_3}}}
+
\frac{1}{C_{\mathrm{d,ch}}}
+
\frac{1}{C_{\mathrm{vdW}}},
\]
where $C_Q$ represents the effective electronic-compressibility response associated with the buried electronic boundary. At the level of this reduced electrostatic model, the defining change is therefore $C_{\mathrm{d,g}}\rightarrow C_Q$. This notation describes a change in the physical origin, location and electronic character of the gate-side series response; it does not imply that the series contribution is eliminated or that $C_Q$ necessarily exceeds $C_{\mathrm{d,g}}$.

Finite $C_Q$ contributes in series with $C_{\mathrm{BiF_3}}$, lowering the nominal TIHGS capacitance. In the representative transistor comparison, the TIHGS- and BiF\textsubscript{3}-only-gated devices have closely matched insulating-layer thicknesses and a common BiF\textsubscript{3}/MoS\textsubscript{2} channel-side material interface. The TIHGS combines lower subthreshold swing ($SS$) and smaller hysteresis with substantially weaker subthreshold drain-bias dependence. Constant-current analysis of the forward sweeps at the common width-normalized drain current $I_{\mathrm{D}}=10$~pA~$\mu$m$^{-1}$ gives signed drain-induced barrier lowering ($DIBL$) coefficients of approximately $+3$ and $-41$~mV~V$^{-1}$ for the TIHGS and BiF\textsubscript{3}-only devices, respectively (Fig.~4f; Methods). The smaller $|DIBL|$ of the TIHGS accompanies improved switching despite its lower nominal capacitance, supporting screening-boundary identity and location as gate-stack design variables complementary to dielectric thickness and permittivity.

Screening-boundary engineering is complementary to, rather than a substitute for, semiconductor-side interface engineering. Channel-side trapping, fixed charge, dielectric disorder and remote-polar-phonon coupling remain relevant to transistor performance~\cite{shin20252d,yoon2025enabling,jung2026advances}. The present device comparison retains the same BiF\textsubscript{3}/MoS\textsubscript{2} channel-side material interface while changing the gate-side boundary condition, thereby allowing the functional consequence of screening-boundary identity and location to be tested separately from a simple change in dielectric thickness or nominal capacitance.

\clearpage

\refstepcounter{suppnote}
\subsection*{Supplementary Note~\thesuppnote. Fluorine Penetration and Self-Limited Surface Conversion}
\label{SN:SRIM}

SF\textsubscript{6}-based processing of Bi\textsubscript{2}Se\textsubscript{3} is known to produce a fluorinated, BiF\textsubscript{3}-rich insulating surface layer~\cite{barton2019impact}. To examine whether fluorine penetration can provide a characteristic length scale for the experimentally observed self-limited surface conversion, we performed stopping and range of ions in matter (SRIM) simulations~\cite{ziegler2010srim}. SRIM treats energetic-ion transport through binary-collision physics, including nuclear and electronic stopping, and provides the depth distribution, projected range, and range straggling of implanted ions. Although SRIM does not reproduce the chemical conversion of Bi\textsubscript{2}Se\textsubscript{3} into BiF\textsubscript{3}, it provides a useful kinematic estimate of the depth over which energetic fluorine can be delivered into Bi\textsubscript{2}Se\textsubscript{3}. Comparison of this ballistic penetration scale with the experimentally observed conversion depth therefore provides a qualitative test of whether fluorine delivery can contribute to defining the spatial extent of the reaction front.

The ion-energy distribution incident on a surface in a radio-frequency (RF)-driven capacitively coupled plasma (CCP) is generally not represented by a single energy. RF sheath modulation, electrode asymmetry and self-bias, collisional transport through the sheath, and charge-exchange processes can broaden and reshape the ion-energy distribution at the sample surface~\cite{lieberman1994principles,kawamura1999ion,gahan2012ion}. In electronegative SF\textsubscript{6}-containing plasmas, negative-ion formation can additionally modify the plasma potential and sheath structure~\cite{franklin2002electronegative,kono2002negative}. Furthermore, as the electrically insulating fluoride layer develops during conversion, surface charging and evolution of the reacted layer can modify the local sheath and subsequent ion-bombardment conditions. Because the ion-energy distribution was not measured directly in our plasma system, we do not assign individual SRIM energies to specific experimental ion populations.

We therefore calculated F\textsuperscript{+} depth distributions at incident energies of 0.5, 1, 2.5, and 5~keV to examine how the accessible ballistic penetration length scale evolves with increasing ion energy. The 5-keV calculation represents a deliberately high-energy upper-bound case rather than an assertion that 5-keV F\textsuperscript{+} ions constitute the dominant ion population under the experimental plasma conditions. Although the experiments employed an SF\textsubscript{6}/Ar plasma containing multiple ionic and neutral reactive species, the simulations were simplified to F\textsuperscript{+} projectiles to isolate the ballistic contribution associated specifically with fluorine delivery. Neutral fluorine radicals, chemical transport, and subsequent solid-state diffusion are not represented by this calculation.

The calculated fluorine depth distributions broaden systematically with increasing incident energy (Extended Data Fig.~\ref{Fig:TEM_SRIM_EDS}). At lower energies, implanted F remains concentrated in the near-surface region, whereas the 5-keV distribution develops an extended tail approaching the $\sim$40-nm depth scale of the experimentally observed saturated converted layer (Extended Data Fig.~\ref{Fig:TEM_SRIM_EDS}). The correspondence between the high-energy ballistic penetration envelope and the independently measured saturation thickness provides an important qualitative clue to the origin of the self-limited conversion depth. Specifically, it shows that a tens-of-nanometres reaction depth is physically accessible to energetic fluorine while suggesting that finite fluorine penetration can help define the maximum spatial extent of the conversion front.

This interpretation is also consistent with the experimentally observed evolution of the converted layer. During the initial stages of SF\textsubscript{6} exposure, fluorine can access and react with the near-surface Bi\textsubscript{2}Se\textsubscript{3}, rapidly increasing the converted-layer thickness. As an insulating BiF\textsubscript{3} layer progressively develops, fluorine delivery to the underlying unconverted Bi\textsubscript{2}Se\textsubscript{3} is expected to become increasingly restricted, while the remaining accessible ballistic and chemically assisted fluorine flux decreases with depth. The combination of a finite fluorine-penetration window and progressive transport limitation through the reacted layer therefore provides a physically plausible explanation for the rapid initial conversion followed by saturation at a characteristic thickness of approximately 40--43~nm.

SRIM itself does not provide a quantitative model of this coupled process. It does not include chemical reactions or BiF\textsubscript{3} phase formation, transport of neutral reactive species, fluorine diffusion, sputter erosion, dynamic changes in target composition and density, plasma-sheath evolution, surface charging, or charge-exchange processes. These mechanisms can modify both fluorine delivery and movement of the reaction front. Nevertheless, the overlap between the calculated high-energy ballistic penetration envelope and the experimentally observed saturation scale is consistent with fluorine penetration contributing to the characteristic conversion depth, while not establishing penetration as the sole or rate-limiting mechanism of the self-limited fluorination process.

\clearpage

\refstepcounter{suppnote}
\subsection*{Supplementary Note~\thesuppnote. Electronic Structure of the Buried Heterophase Boundary}
\label{SN:Buried_Electronic_Boundary}

In the pristine Bi\textsubscript{2}Se\textsubscript{3} slab comprising six quintuple layers (QLs), the position-resolved electronic-structure calculations in Fig.~\ref{Fig:Buried_Boundary} were used to establish the reference electronic boundary before surface conversion. Bi\textsubscript{2}Se\textsubscript{3} is a prototypical three-dimensional topological insulator whose bulk topology gives rise to a Dirac-like surface state within the bulk band gap~\cite{chen2009experimental,moore2010birth,pesin2012spintronics}. Consistent with this established picture, the pristine slab exhibits a Dirac-like surface state within the calculated 0.31-eV bulk gap, with its Dirac point ($E_{\mathrm{D}}$) approximately 0.043~eV below the Fermi level ($E_{\mathrm{F}}$) (Fig.~\ref{Fig:Buried_Boundary}a,b).

Formation of the a-BiF\textsubscript{3}/c-Bi\textsubscript{2}Se\textsubscript{3} heterophase boundary produces a strongly position-dependent electronic reconstruction. The electronic states projected onto the immediate heterophase interface are gap-opened by approximately 0.72~eV, whereas the adjacent c-Bi\textsubscript{2}Se\textsubscript{3} subinterface QL develops a reconstructed gap-closed Bi\textsubscript{2}Se\textsubscript{3}-derived Dirac-like feature centred approximately 0.146~eV below $E_{\mathrm{F}}$, together with finite spectral weight extending to the Fermi level (Fig.~\ref{Fig:Buried_Boundary}c,d). The surface QL at the opposite side of the slab retains a pristine-like Dirac dispersion. Thus, the $\sim0.146$-eV value specifies the energetic position of the reconstructed subinterface feature relative to $E_{\mathrm{F}}$ and is not an interfacial band gap. More importantly, the calculations show that the chemical heterophase interface and the electronically active boundary are spatially distinct: surface conversion opens the immediate interface while relocating a gap-closed Bi\textsubscript{2}Se\textsubscript{3}-derived electronic response into the adjacent subinterface layer.

The calculated dipole-moment profile across the a-BiF\textsubscript{3}/c-Bi\textsubscript{2}Se\textsubscript{3} structure is sharply localized near the heterophase boundary (Fig.~\ref{Fig:Buried_Boundary}e). This localization is consistent with interfacial charge redistribution and with a local electrostatic contribution to the energetic reconstruction of the Bi\textsubscript{2}Se\textsubscript{3}-derived state. We do not assign the shift of the Dirac-like feature from $\sim0.043$ to $\sim0.146$~eV uniquely to the interface dipole, because local bonding, chemical reconstruction and band bending are not separately decomposed. Rather, the interface-localized dipole and the position-resolved band reconstruction provide complementary microscopic evidence that the electronic boundary is reorganized by formation of a-BiF\textsubscript{3}. This physical picture is consistent with previous reports that surface modification or etching of Bi\textsubscript{2}Se\textsubscript{3} can displace or reconstruct its surface-state distribution~\cite{barton2019impact,yue2024topological}.

A finite electronic density of states at the Fermi level provides a microscopic basis for finite electronic compressibility and quantum capacitance,
\[
C_Q=e^{2}D(E_{\mathrm{F}})
\]
in the zero-temperature limit. The calculated near-$E_{\mathrm{F}}$ spectral weight of the subinterface boundary is therefore consistent with the finite $C_Q$ extracted independently from the TIHGS capacitance measurements. The relevance of this response is not that $C_Q$ increases the total gate capacitance---it does not. Rather, finite electronic compressibility allows the heterophase-induced buried boundary to support screening charge while the underlying c-Bi\textsubscript{2}Se\textsubscript{3} remains electrically coupled to the external gate electrode. In this sense, $C_Q$ is treated as an electrostatic signature of the relocated electronically active boundary rather than as an intrinsic dielectric contribution of BiF\textsubscript{3} or a capacitance-enhancement mechanism.

The calculations do not constitute an independent determination of the topological invariant of the chemically reconstructed interface. We do not extract an interface-specific $Z_2$ invariant or explicitly calculate the spin texture of the reconstructed subinterface state. We therefore interpret the calculations conservatively as evidence for a reconstructed Bi\textsubscript{2}Se\textsubscript{3}-derived Dirac-like electronic boundary whose relation to topological-insulator bulk--boundary physics is motivated by the established bulk topology of Bi\textsubscript{2}Se\textsubscript{3}. Additional chemically induced, band-bending-related or reconstructed states may coexist and contribute to the experimentally extracted electronic compressibility.

\clearpage

\refstepcounter{suppnote}
\subsection*{Supplementary Note~\thesuppnote. Effective Quantum-Capacitance Extraction and Model Assumptions}
\label{SN:Quantum_Capacitance}

The effective quantum-capacitance contribution of the buried BiF\textsubscript{3}/Bi\textsubscript{2}Se\textsubscript{3} boundary was estimated by comparing the measured TIHGS capacitance with the dielectric capacitance expected for BiF\textsubscript{3}. Throughout this analysis, all capacitances are expressed per unit area. The TIHGS response was represented as
\[
\frac{1}{C_{\mathrm{TIHGS}}}
=
\frac{1}{C_Q}
+
\frac{1}{C_{\mathrm{BiF_3}}},
\]
where
\[
C_{\mathrm{BiF_3}}
=
\frac{\varepsilon_{0}\varepsilon_{\mathrm{BiF_3}}}
{t_{\mathrm{BiF_3}}}.
\]
Here $C_Q$ denotes an effective electronic-compressibility contribution associated with the buried electronically active boundary. It is a finite series response and should not be interpreted as an intrinsic dielectric property of BiF\textsubscript{3}, a universal material constant, or a capacitance-enhancement term.

The BiF\textsubscript{3} dielectric response used for this analysis was obtained from the fully converted BiF\textsubscript{3}-only (p20) reference. This structure was formed from a nominally 20-nm-thick polycrystalline Bi\textsubscript{2}Se\textsubscript{3} film and exhibited an independently measured post-conversion BiF\textsubscript{3} thickness of $\sim$23~nm. Using this physical thickness, the measured capacitor statistics yield a mean relative permittivity $\varepsilon_{\mathrm{BiF_3}}=28.65$ and a median value of 30.16. The mean value was used as the reference BiF\textsubscript{3} dielectric response for the primary $C_Q$ extraction below. This extraction assumes that the converted BiF\textsubscript{3} region within the TIHGS has the same dielectric response as the fully converted BiF\textsubscript{3}-only reference. Any systematic difference in the local structure, density, composition, or dielectric response of BiF\textsubscript{3} between the partially and fully converted configurations would renormalize the extracted $C_Q$. Accordingly, the reported $C_Q$ values should be regarded as effective interfacial quantum-capacitance responses within this reference-based series-capacitance model rather than as model-independent microscopic constants.

For the TIHGS-13 (p20) structure,
\[
C_{\mathrm{TIHGS}}
=
1.10~\mu\mathrm{F\,cm^{-2}}
\]
and, using $t_{\mathrm{BiF_3}}\approx13$~nm,
\[
C_{\mathrm{BiF_3}}
=
1.95~\mu\mathrm{F\,cm^{-2}},
\]
yielding
\[
C_Q
=
\left(
\frac{1}{C_{\mathrm{TIHGS}}}
-
\frac{1}{C_{\mathrm{BiF_3}}}
\right)^{-1}
\approx
2.52~\mu\mathrm{F\,cm^{-2}}.
\]

For the TIHGS-43 (p50) structure,
\[
C_{\mathrm{TIHGS}}
=
0.305~\mu\mathrm{F\,cm^{-2}}
\]
and, using $t_{\mathrm{BiF_3}}\approx43$~nm,
\[
C_{\mathrm{BiF_3}}
=
0.590~\mu\mathrm{F\,cm^{-2}},
\]
giving
\[
C_Q
\approx
0.631~\mu\mathrm{F\,cm^{-2}}.
\]

As a sensitivity check, using the median BiF\textsubscript{3} relative permittivity of 30.16 instead of the mean value gives $C_Q\approx2.37~\mu\mathrm{F\,cm^{-2}}$ for TIHGS-13 (p20) and $C_Q\approx0.599~\mu\mathrm{F\,cm^{-2}}$ for TIHGS-43 (p50). Thus, the inference of a finite interfacial electronic-compressibility contribution is not sensitive to whether the mean or median BiF\textsubscript{3} reference permittivity is used.

The finite $C_Q$ explains why the measured TIHGS capacitance is lower than that expected from BiF\textsubscript{3} alone. Accordingly, the present MIM measurements do not demonstrate a net capacitance or $EOT$ advantage of the TIHGS structure over the BiF\textsubscript{3}-only reference. Nor do they independently determine the magnitude of the conventional metal/BiF\textsubscript{3} gate-side interfacial capacitance, $C_{\mathrm{d,g}}$, and therefore they do not establish that $C_Q>C_{\mathrm{d,g}}$. Within this reference-based series-capacitance model, the capacitance measurements instead require a finite additional series response associated with the buried heterophase boundary.

The distinction between $C_Q$ and a conventional gate-side interfacial or dead-layer capacitance is therefore primarily one of physical origin and boundary identity rather than the existence versus absence of a finite series term. Finite gate-electrode/dielectric interfacial responses are well established as non-scaling contributions to deeply scaled gate-stack electrostatics~\cite{lau2023dielectrics,pourfath2026device}. Conventional gate-side contributions can arise from finite electrode screening, altered interfacial polarizability, bonding, chemical reconstruction, or structural disorder. In the TIHGS architecture, the metal electrode is electrically coupled to residual Bi\textsubscript{2}Se\textsubscript{3}, so the conventional metal/BiF\textsubscript{3} interface no longer defines the active gate-side screening boundary. The remaining $C_Q$ instead characterizes the effective electronic compressibility of the deliberately formed BiF\textsubscript{3}/Bi\textsubscript{2}Se\textsubscript{3} boundary. Thus, ``boundary relocation'' rather than ``elimination of all interfacial capacitance'' is the appropriate electrostatic description. The transistor comparison in Fig.~\ref{Fig:FET} separately evaluates whether changing this boundary condition has a functional consequence for gate control.

The differing $C_Q$ values obtained for the two conversion geometries indicate that the extracted quantity depends on the microscopic electronic structure and electrostatic environment of the reconstructed boundary rather than representing a universal material parameter. This interpretation is consistent with the position-resolved electronic-structure calculation in Fig.~\ref{Fig:Buried_Boundary}d, which predicts finite Bi\textsubscript{2}Se\textsubscript{3}-derived spectral weight at the Fermi level near the buried heterophase boundary.

An independent constraint on the spatial origin of this additional series-capacitance response is provided by the crystalline TIHGS-43 (c$N$) structures. When the nominal starting crystalline Bi\textsubscript{2}Se\textsubscript{3} thickness is varied from $\sim$49 to $\sim$179~nm while maintaining approximately the same converted BiF\textsubscript{3} thickness, $\varepsilon_{\mathrm{eff,TIHGS}}$ remains within $\sim$15.9--17.4 without a systematic thickness dependence (Extended Data Fig.~\ref{Fig:MIM_Ref}f). This weak dependence argues against a dominant electrostatic contribution from the Bi\textsubscript{2}Se\textsubscript{3} bulk and is instead consistent with a contribution localized near the BiF\textsubscript{3}/Bi\textsubscript{2}Se\textsubscript{3} boundary.

Capacitance measurements alone cannot uniquely distinguish the reconstructed Bi\textsubscript{2}Se\textsubscript{3}-derived boundary state from additional electronic states that may coexist at a band-bent or chemically reconstructed interface. We therefore interpret $C_Q$ conservatively as an effective electronic-compressibility response associated with the buried electronically active boundary. Its finite magnitude, together with the calculated near-Fermi spectral weight, provides mutually consistent electrostatic and microscopic evidence for an electronically compressible buried boundary.

\clearpage

\refstepcounter{suppnote}
\subsection*{Supplementary Note~\thesuppnote. Field-Assisted Leakage, Band Alignment and Effective Tunnelling Barrier}
\label{SN:Emission_Tunnelling}

To examine the field dependence of gate-stack leakage in the TIHGS MIM capacitors, we analysed the current-density--voltage characteristics using field-assisted emission and tunnelling representations, including Simmons-modified Schottky emission, Poole--Frenkel emission, and Fowler--Nordheim tunnelling (Extended Data Fig.~\ref{Fig:TIHGS_Band}a--c)~\cite{simmons1965richardson,lenzlinger1969fowler,chiu2014conduction}. The electric field was calculated as
\[
E=\frac{|V|}{t_{\mathrm{BiF_3}}},
\]
where $t_{\mathrm{BiF_3}}$ is the physical thickness of the insulating BiF\textsubscript{3} layer. The electrically conducting Bi\textsubscript{2}Se\textsubscript{3} region was not included in the dielectric thickness used for this field conversion. This treatment assumes that the applied voltage drop relevant to high-field transport occurs predominantly across the insulating BiF\textsubscript{3} layer; any additional voltage drop associated with the interfaces is not independently resolved and is therefore incorporated into the effective transport parameters extracted below.

At intermediate electric fields, the data in Extended Data Fig.~\ref{Fig:TIHGS_Band}b exhibit approximately linear behaviour when represented as $\ln(J/E)$ versus $E^{1/2}$, consistent with Simmons-modified Schottky and Poole--Frenkel descriptions~\cite{simmons1965richardson,chiu2014conduction}. The corresponding field-lowering coefficients are
\[
\beta_{\mathrm{S}}
=
\left[
\frac{q^{3}}
{4\pi\varepsilon_{0}\varepsilon_{\mathrm{r}}(\infty)}
\right]^{1/2}
\]
for Schottky-type barrier lowering and
\[
\beta_{\mathrm{PF}}
=
\left[
\frac{q^{3}}
{\pi\varepsilon_{0}\varepsilon_{\mathrm{r}}(\infty)}
\right]^{1/2}
=
2\beta_{\mathrm{S}}
\]
for Poole--Frenkel emission, where $q$ is the elementary charge, $\varepsilon_{0}$ is the vacuum permittivity, and $\varepsilon_{\mathrm{r}}(\infty)$ is the ion-clamped high-frequency relative permittivity of the dielectric.

The Simmons-modified Schottky and Poole--Frenkel representations yield $\varepsilon_{\mathrm{r}}(\infty)\approx4.99$ and $\approx4.89$, respectively, corresponding to refractive indices $n=\sqrt{\varepsilon_{\mathrm{r}}(\infty)}$ of $\sim$2.23 and $\sim$2.21. These values are reasonably close to the first-principles values calculated for amorphous BiF\textsubscript{3}, $\varepsilon_{\mathrm{r}}(\infty)=4.056$ and $n=2.014$, providing a consistency check between the field-dependent transport response and the calculated electronic polarizability.

At higher electric fields, the Fowler--Nordheim (F--N) representation becomes dominant~\cite{lenzlinger1969fowler,chiu2014conduction}. Within the conventional triangular-barrier approximation, the F--N current density can be expressed in the form
\[
J_{\mathrm{FN}}
\propto
E^{2}
\exp
\left[
-
\frac{
4\sqrt{2m_{\mathrm{t}}^{*}}\,
\Phi_{\mathrm{B,FN}}^{3/2}
}{
3q\hbar E
}
\right],
\]
where $m_{\mathrm{t}}^{*}$ is the tunnelling effective mass and $\Phi_{\mathrm{B,FN}}$ is the effective electron-injection barrier height. In this expression, $\Phi_{\mathrm{B,FN}}$ is expressed in joules. The corresponding linearized form is
\[
\ln\left(\frac{J}{E^{2}}\right)
=
A_{\mathrm{FN}}
-
\frac{
4\sqrt{2m_{\mathrm{t}}^{*}}\,
\Phi_{\mathrm{B,FN}}^{3/2}
}{
3q\hbar
}
\frac{1}{E},
\]
where $A_{\mathrm{FN}}$ contains the field-independent prefactor.

Because the BiF\textsubscript{3} thickness is fixed for a given MIM capacitor and $E=|V|/t_{\mathrm{BiF_3}}$, the equivalent voltage-domain representation used in Extended Data Fig.~\ref{Fig:TIHGS_Band}c is
\[
\ln\left(\frac{J}{V^{2}}\right)
=
A_{\mathrm{FN}}^{\prime}
-
\frac{
4t_{\mathrm{BiF_3}}
\sqrt{2m_{\mathrm{t}}^{*}}\,
\Phi_{\mathrm{B,FN}}^{3/2}
}{
3q\hbar
}
\frac{1}{|V|}.
\]
For a fixed capacitor area, replacing current density $J$ with total current $I$ changes only the intercept and not the slope. Accordingly, if $S_{\mathrm{FN}}$ denotes the slope of the linear high-field region in a plot of $\ln(J/V^{2})$ versus $1/|V|$, the effective F--N barrier height is obtained from
\[
\Phi_{\mathrm{B,FN}}
=
\left[
\frac{
3q\hbar |S_{\mathrm{FN}}|
}{
4t_{\mathrm{BiF_3}}\sqrt{2m_{\mathrm{t}}^{*}}
}
\right]^{2/3}.
\]
The quantity obtained from this equation is an energy in joules and was divided by $q$ to report it in electronvolts.

For the Au/BiF\textsubscript{3}/Bi\textsubscript{2}Se\textsubscript{3}/Au/Ti TIHGS MIM capacitor analysed in Extended Data Fig.~\ref{Fig:TIHGS_Band}a--c, $t_{\mathrm{BiF_3}}\approx43$~nm. We adopted $m_{\mathrm{t}}^{*}=0.25m_{0}$ as a representative tunnelling mass within the range reported for high-$\kappa$ dielectrics, where $m_{0}$ is the free-electron mass~\cite{li2004selected}. Using the slope of the high-field linear regime gives an effective F--N barrier height of
\[
\Phi_{\mathrm{B,FN}}
\approx
0.27~\mathrm{eV}.
\]

The high-field transport analysis must be distinguished from the equilibrium microscopic electronic structure of the heterophase boundary. Figure~\ref{Fig:Buried_Boundary}c--e shows that the immediate a-BiF\textsubscript{3}/c-Bi\textsubscript{2}Se\textsubscript{3} interface is gap-opened, whereas a reconstructed gap-closed Bi\textsubscript{2}Se\textsubscript{3}-derived state resides in the adjacent c-Bi\textsubscript{2}Se\textsubscript{3} subinterface QL and the calculated dipole-moment profile is strongly localized near the heterophase boundary. The localized dipole establishes interfacial charge redistribution but does not by itself determine the magnitude of the reconstructed-state energy shift. The band alignment of the heterophase therefore cannot, in general, be represented solely by rigid alignment of isolated-material electron affinity and work function.

Extended Data Fig.~\ref{Fig:TIHGS_Band}d shows a vacuum-referenced constituent alignment before Au contact. Using the calculated Bi\textsubscript{2}Se\textsubscript{3} work function $\phi_{\mathrm{Bi_2Se_3}}=5.15$~eV and amorphous-BiF\textsubscript{3} electron affinity $\chi_{\mathrm{BiF_3}}=4.3$~eV gives a nominal separation of approximately 0.85~eV between the Bi\textsubscript{2}Se\textsubscript{3} Fermi level and the BiF\textsubscript{3} conduction-band reference. This value is a vacuum-level reference construction rather than an exact equilibrium conduction-band offset, because the interface dipole, chemical reconstruction and associated band bending can modify the local electrostatic potential. The reconstructed Bi\textsubscript{2}Se\textsubscript{3}-derived subinterface feature identified in Fig.~\ref{Fig:Buried_Boundary}d is therefore shown separately at approximately 0.146~eV below $E_{\mathrm{F}}$.

After Au contact, the electrochemical potentials equilibrate across the contacted stack (Extended Data Fig.~\ref{Fig:TIHGS_Band}e). The Au work function, $\phi_{\mathrm{Au}}=5.1$~eV, differs from the calculated Bi\textsubscript{2}Se\textsubscript{3} work function by only approximately 0.05~eV, whereas the corresponding vacuum-referenced Au/BiF\textsubscript{3} separation is approximately 0.8~eV. These quantities provide a reference energetic sequence for the contacted capacitor but do not imply that the local heterophase or metal/dielectric band offsets are independently determined with this precision. The interface-localized dipole identified in Fig.~\ref{Fig:Buried_Boundary}e and any residual band bending remain incorporated implicitly in the contacted-state alignment.

Under the positive high-field condition used for the Fowler--Nordheim analysis, the potential drop across BiF\textsubscript{3} tilts the insulating barrier and enables tunnelling from the electrically connected residual Bi\textsubscript{2}Se\textsubscript{3} towards accessible BiF\textsubscript{3} conduction-band states (Extended Data Fig.~\ref{Fig:TIHGS_Band}f). The experimentally extracted $\Phi_{\mathrm{B,FN}}\approx0.27$~eV is therefore assigned to an effective high-field Bi\textsubscript{2}Se\textsubscript{3}-to-BiF\textsubscript{3} injection barrier. Its difference from the simple vacuum-referenced equilibrium separation is not interpreted as a direct measurement of the interface dipole or band bending; rather, it reflects a field-modified tunnelling barrier in a chemically and electronically reconstructed heterophase system. Because $\Phi_{\mathrm{B,FN}}\propto(m_{\mathrm{t}}^{*})^{-1/3}$ and the BiF\textsubscript{3}-specific tunnelling mass was not independently measured, the extracted value remains a model-dependent effective barrier rather than an exact equilibrium band offset.

Finally, the Fowler--Nordheim regime occurs well outside the gate-voltage window used for the MoS\textsubscript{2} transistor measurements and therefore does not represent the dominant transport process during normal transistor operation. Linearized Schottky, Poole--Frenkel and Fowler--Nordheim representations are also not unique microscopic fingerprints of individual leakage mechanisms~\cite{chiu2014conduction}. We therefore use these analyses as mutually consistent descriptions of field-dependent transport and as a means to estimate the effective high-field injection barrier, rather than as independent determinations of the equilibrium interface band structure.

\clearpage

\refstepcounter{suppnote}
\subsection*{Supplementary Note~\thesuppnote. Flat-Band Extraction and Semiconductor Screening}
\label{SN:Flat-Band_Screening}

The flat-band response and semiconductor-side screening parameters were analysed using the frequency-dependent capacitance--voltage ($C$--$V$) characteristics of the Au/Ag/Pt/WSe\textsubscript{2}/BiF\textsubscript{3}/Bi\textsubscript{2}Se\textsubscript{3}/Au/Ti metal--insulator--semiconductor capacitor (MISCAP). Throughout this analysis, all capacitances are expressed per unit area.

The flat-band voltage, $V_{\mathrm{FB}}$, was determined from the accumulation-to-depletion transition using the capacitance--voltage inflection-point method identified from the second derivative of the measured $C$--$V$ characteristic~\cite{winter2013new}. The 10-kHz $C$--$V$ characteristic was used because the corresponding transition remained within the experimentally accessible low-leakage bias window. At higher frequencies, the transition shifted toward gate voltages above the measurement range, where increasing dielectric leakage prevented reliable determination of the flat-band condition. The physically relevant flat-band voltage was identified from the principal accumulation-to-depletion inflection in the 10-kHz characteristic, corresponding to

\begin{equation}
\left.
\frac{\mathrm{d}^{2}C_{\mathrm{m}}}
{\mathrm{d}V_{\mathrm{G}}^{2}}
\right|_{V_{\mathrm{G}}=V_{\mathrm{FB}}}
=
0,
\tag{S1}
\end{equation}

with a corresponding change in the sign of $\mathrm{d}^{2}C_{\mathrm{m}}/\mathrm{d}V_{\mathrm{G}}^{2}$. This analysis gives

\begin{equation}
V_{\mathrm{FB}}
\approx
0.58~\mathrm{V},
\tag{S2}
\end{equation}

as shown in Extended Data Fig.~\ref{Fig:MISCAP}c. The measured 10-kHz capacitance at the flat-band condition is

\begin{equation}
C_{\mathrm{FB}}
=
C_{\mathrm{m}}(V_{\mathrm{FB}})
=
2.34\times10^{-7}~\mathrm{F\,cm^{-2}}.
\tag{S3}
\end{equation}

We used an independent graphical consistency analysis to determine the effective capacitance of the TIHGS. At the independently determined $V_{\mathrm{FB}}$, the two dimensionless functions

\begin{equation}
f_{1}
=
\frac{1}{
\sqrt{
\displaystyle
\frac{3k_{\mathrm{B}}T}{q}
\frac{1}{C_{\mathrm{m}}}
\frac{\mathrm{d}C_{\mathrm{m}}}{\mathrm{d}V_{\mathrm{G}}}
}
}
-1
\tag{S4}
\end{equation}

and

\begin{equation}
f_{2}
=
\frac{C_{\mathrm{m}}}
{C_{\mathrm{g}}-C_{\mathrm{m}}}
\tag{S5}
\end{equation}

were evaluated, where $k_{\mathrm{B}}$ is the Boltzmann constant, $T$ is the absolute temperature, $q$ is the elementary charge, and $C_{\mathrm{g}}$ is the effective capacitance of the complete TIHGS. Trial values of $C_{\mathrm{g}}$ were varied until $f_{1}=f_{2}$ at the independently determined $V_{\mathrm{FB}}$ (Extended Data Fig.~\ref{Fig:MISCAP}d). The resulting gate-stack response corresponds to

\begin{equation}
\varepsilon_{\mathrm{eff,TIHGS}}
\approx
16.
\tag{S6}
\end{equation}

For the approximately 43-nm-thick converted BiF\textsubscript{3} layer in the MISCAP, this effective relative permittivity gives

\begin{equation}
C_{\mathrm{g}}
=
\frac{
\varepsilon_{0}\varepsilon_{\mathrm{eff,TIHGS}}
}{
t_{\mathrm{BiF_3}}
}
=
3.29\times10^{-7}~\mathrm{F\,cm^{-2}},
\tag{S7}
\end{equation}

where $\varepsilon_{0}$ is the vacuum permittivity. As in the TIHGS MIM analysis, $\varepsilon_{\mathrm{eff,TIHGS}}$ is a BiF\textsubscript{3}-thickness-normalized stack-level electrostatic quantity and should not be interpreted as the intrinsic dielectric constant of BiF\textsubscript{3}. The value $\varepsilon_{\mathrm{eff,TIHGS}}\approx16$ is consistent with the finite boundary response independently observed in the TIHGS MIM capacitors and is therefore not interpreted as evidence for a capacitance or $EOT$ enhancement. Its purpose here is to parameterize the effective TIHGS capacitance required to de-embed the semiconductor-side response.

The measured flat-band capacitance contains the TIHGS capacitance in series with the semiconductor-side response. The effective semiconductor-side flat-band capacitance was therefore obtained from

\begin{equation}
\frac{1}{C_{\mathrm{FB}}}
=
\frac{1}{C_{\mathrm{g}}}
+
\frac{1}{C_{\mathrm{s,FB}}^{*}},
\tag{S8}
\end{equation}

giving

\begin{equation}
C_{\mathrm{s,FB}}^{*}
=
\left(
\frac{1}{C_{\mathrm{FB}}}
-
\frac{1}{C_{\mathrm{g}}}
\right)^{-1}
=
8.08\times10^{-7}~\mathrm{F\,cm^{-2}}.
\tag{S9}
\end{equation}

The corresponding effective screening length was estimated as

\begin{equation}
L_{\mathrm{D,eff}}
=
\frac{
\varepsilon_{0}\varepsilon_{\perp,\mathrm{WSe_2}}
}{
C_{\mathrm{s,FB}}^{*}
},
\tag{S10}
\end{equation}

where $\varepsilon_{\perp,\mathrm{WSe_2}}$ is the out-of-plane relative permittivity of WSe\textsubscript{2}. Using $T=300$~K and a representative value $\varepsilon_{\perp,\mathrm{WSe_2}}=4.2$, as adopted in continuum electrostatic modelling of bulk-like WSe\textsubscript{2} flakes~\cite{yu2017photogenerated}, gives

\begin{equation}
L_{\mathrm{D,eff}}
\approx
4.60~\mathrm{nm}.
\tag{S11}
\end{equation}

The corresponding apparent volumetric majority-carrier concentration was estimated from

\begin{equation}
N_{\mathrm{app}}
=
\frac{
k_{\mathrm{B}}T
\varepsilon_{0}\varepsilon_{\perp,\mathrm{WSe_2}}
}{
q^{2}L_{\mathrm{D,eff}}^{2}
},
\tag{S12}
\end{equation}

yielding

\begin{equation}
N_{\mathrm{app}}
\approx
2.83\times10^{17}~\mathrm{cm^{-3}}.
\tag{S13}
\end{equation}

Because the flat-band response is evaluated at 10~kHz, traps that remain electrically responsive at this frequency can contribute to $C_{\mathrm{s,FB}}^{*}$. Accordingly, $L_{\mathrm{D,eff}}$ and $N_{\mathrm{app}}$ are interpreted as effective low-frequency screening parameters rather than intrinsic bulk-semiconductor quantities. The measured WSe$_2$ thickness, $t_{\mathrm{WSe_2}}=37.37$~nm, nevertheless substantially exceeds $2L_{\mathrm{D,eff}}\approx9.20$~nm, indicating that electrostatic perturbations originating from the two opposing WSe$_2$ surfaces do not strongly overlap across the full semiconductor thickness.

\clearpage

\refstepcounter{suppnote}
\subsection*{Supplementary Note~\thesuppnote. Effective Interface-Trap Extraction and Limitations}
\label{SN:Interface-Trap}

The effective electrically active trap density, $D_{\mathrm{it,eff}}$, was estimated from the frequency-dependent capacitance--voltage characteristics using the high--low-frequency Castagn\'e--Vapaille approach~\cite{castagne1971interface,engelherbert2010interface}. The method exploits the different frequency response of electrically active interface and near-interfacial traps and has been widely used to estimate trap densities from metal--insulator--semiconductor capacitance measurements, although the extracted values can depend on the assumed semiconductor response, measurement frequency window, and gate-stack electrostatics~\cite{engelherbert2010interface}. In this analysis, we denote the measured capacitances at 10~kHz and 1~MHz as $C_{\mathrm{LF}}$ and $C_{\mathrm{HF}}$, respectively, so traps that respond differently over this frequency range contribute to the measured capacitance dispersion.

Throughout this analysis, all capacitances are expressed per unit area. The measured MISCAP capacitance contains the TIHGS electrostatic element in series with the semiconductor-side response. We use $C_{\mathrm{g}}\equiv C_{\mathrm{TIHGS}}=3.29\times10^{-7}$~F~cm$^{-2}$, obtained from the 10-kHz flat-band consistency analysis in Supplementary Note~\ref{SN:Flat-Band_Screening}. Because an independently resolved frequency-dependent $C_{\mathrm{g}}$ is unavailable, this value is used for both the 10-kHz and 1-MHz de-embedding; any residual TIHGS dispersion is therefore included in the effective quantity $D_{\mathrm{it,eff}}$. The effective semiconductor-side capacitances were calculated as

\begin{equation}
C_{\mathrm{s,LF}}^{*}
=
\left(
\frac{1}{C_{\mathrm{LF}}}
-
\frac{1}{C_{\mathrm{g}}}
\right)^{-1},
\tag{S14}
\end{equation}

and

\begin{equation}
C_{\mathrm{s,HF}}^{*}
=
\left(
\frac{1}{C_{\mathrm{HF}}}
-
\frac{1}{C_{\mathrm{g}}}
\right)^{-1},
\tag{S15}
\end{equation}

where $C_{\mathrm{LF}}$ and $C_{\mathrm{HF}}$ are the measured capacitances at 10~kHz and 1~MHz, respectively.

Within the conventional high--low-frequency approximation, the low-frequency semiconductor-side response includes semiconductor charge modulation and trap contributions that remain responsive at lower frequencies, whereas the corresponding trap response is reduced at higher frequencies. The frequency-dependent excess capacitance was therefore defined as

\begin{equation}
C_{\mathrm{it,eff}}
=
C_{\mathrm{s,LF}}^{*}
-
C_{\mathrm{s,HF}}^{*},
\tag{S16}
\end{equation}

and the corresponding effective electrically active trap density as

\begin{equation}
D_{\mathrm{it,eff}}
=
\frac{
C_{\mathrm{it,eff}}
}{q}
=
\frac{1}{q}
\left[
\left(
\frac{1}{C_{\mathrm{LF}}}
-
\frac{1}{C_{\mathrm{g}}}
\right)^{-1}
-
\left(
\frac{1}{C_{\mathrm{HF}}}
-
\frac{1}{C_{\mathrm{g}}}
\right)^{-1}
\right].
\tag{S17}
\end{equation}

The quantitative interpretation focused on the bias-dependent depletion-to-accumulation transition. The nearly gate-voltage-independent depletion floor was excluded because the semiconductor charge response is constrained by the finite WSe$_2$ thickness. Strong-accumulation values were likewise excluded because $C_{\mathrm{LF}}$ approaches $C_{\mathrm{g}}$, making the semiconductor-side capacitance increasingly sensitive to small experimental uncertainties. Within the range (Extended Data Fig.~\ref{Fig:MISCAP}e), the minimum extracted value is

\begin{equation}
D_{\mathrm{it,eff}}^{\mathrm{min}}
\approx
7.48\times10^{11}
~\mathrm{cm^{-2}\,eV^{-1}}.
\tag{S18}
\end{equation}

The parallel-conductance response, $G_{\mathrm{p}}/\omega$, was also examined over the same gate-voltage and frequency ranges (Extended Data Fig.~7b). No well-resolved $G_{\mathrm{p}}/\omega$ maximum suitable for conventional conductance-method extraction of trap density or characteristic time constant was observed within the accessible gate-bias range, while higher positive gate voltages were precluded by increasing dielectric leakage. The effective trap density was therefore evaluated using the high--low-frequency capacitance analysis described above.

The high--low-frequency method attributes the low--high capacitance difference to electrically active trap-related dispersion within the selected frequency window. In the WSe$_2$/TIHGS MISCAP, this response may include contributions from the WSe$_2$/BiF$_3$ interface, near-interfacial or border traps, residual TIHGS dispersion, distributed WSe$_2$ resistance, contact resistance, dielectric leakage, and measurement parasitics. Moreover, because 10~kHz is not a true quasi-static limit, substantially slower traps are not sampled. Accordingly, the extracted $D_{\mathrm{it,eff}}$ should be interpreted as an effective electrically active trap response within the experimental frequency window rather than as a unique microscopic density of states exclusively at the WSe\textsubscript{2}/BiF\textsubscript{3} interface, consistent with the known model and frequency dependence of high--low-frequency interface-trap extraction~\cite{engelherbert2010interface}.

\clearpage


\bibliography{sn-bibliography}

@article{su2026high,
  title={High-Transconductance Molybdenum Disulfide Top-Gate Transistors Using Epitaxial Interface Engineering},
  author={Su, Yuan-Chun and Mao, Po-Sen and Shih, Chih-Yao and Wang, Shih-Ting and Shen, Yun-Yang and Jian, Zih-Siang and Hu, Hsiang-Chi and Tseng, Wei-Chen and Sung, Hsin-Ya and Lin, Chih-Yen and others},
  journal={Nature Electronics},
  pages={1--10},
  year={2026},
  publisher={Nature Publishing Group UK London}
}

@article{sun2021topological,
  title={Topological Insulator Metamaterial with Giant Circular Photogalvanic Effect},
  author={Sun, Xinxing and Adamo, Giorgio and Eginligil, Mustafa and Krishnamoorthy, Harish NS and Zheludev, Nikolay I and Soci, Cesare},
  journal={Science Advances},
  volume={7},
  number={14},
  pages={eabe5748},
  year={2021},
  publisher={American Association for the Advancement of Science}
}

@article{barton2019impact,
  title={Impact of Etch Processes on the Chemistry and Surface States of the Topological Insulator {Bi$_2$Se$_3$}},
  author={Barton, Adam T and Walsh, Lee A and Smyth, Christopher M and Qin, Xiaoye and Addou, Rafik and Cormier, Christopher and Hurley, Paul K and Wallace, Robert M and Hinkle, Christopher L},
  journal={ACS Applied Materials \& Interfaces},
  volume={11},
  number={35},
  pages={32144--32150},
  year={2019},
  publisher={ACS Publications}
}

@article{lau2023dielectrics,
  title={Dielectrics for Two-Dimensional Transition-Metal Dichalcogenide Applications},
  author={Lau, Chit Siong and Das, Sarthak and Verzhbitskiy, Ivan A and Huang, Ding and Zhang, Yiyu and Talha-Dean, Teymour and Fu, Wei and Venkatakrishnarao, Dasari and Johnson Goh, Kuan Eng},
  journal={ACS Nano},
  volume={17},
  number={11},
  pages={9870--9905},
  year={2023},
  publisher={ACS Publications}
}

@article{li2019uniform,
  title={Uniform and Ultrathin High-$\kappa$ Gate Dielectrics for Two-Dimensional Electronic Devices},
  author={Li, Weisheng and Zhou, Jian and Cai, Songhua and Yu, Zhihao and Zhang, Jialin and Fang, Nan and Li, Taotao and Wu, Yun and Chen, Tangsheng and Xie, Xiaoyu and others},
  journal={Nature Electronics},
  volume={2},
  number={12},
  pages={563--571},
  year={2019},
  publisher={Nature Publishing Group UK London}
}

@article{li2020native,
  title={A Native Oxide High-$\kappa$ Gate Dielectric for Two-Dimensional Electronics},
  author={Li, Tianran and Tu, Teng and Sun, Yuanwei and Fu, Huixia and Yu, Jia and Xing, Lei and Wang, Ziang and Wang, Huimin and Jia, Rundong and Wu, Jinxiong and others},
  journal={Nature Electronics},
  volume={3},
  number={8},
  pages={473--478},
  year={2020},
  publisher={Nature Publishing Group UK London}
}

@article{kim2025gate,
  title={Gate Stack Engineering of Two-Dimensional Transistors},
  author={Kim, Yeon Ho and Lee, Donghun and Huh, Woong and Lee, Jaeho and Lee, Donghyun and Wang, Gunuk and Park, Jaehyun and Ha, Daewon and Lee, Chul-Ho},
  journal={Nature Electronics},
  volume={8},
  pages={770--784},
  year={2025},
  publisher={Nature Publishing Group UK London}
}

@article{xu2023scalable,
  title={Scalable Integration of Hybrid High-$\kappa$ Dielectric Materials on Two-Dimensional Semiconductors},
  author={Xu, Yongshan and Liu, Teng and Liu, Kailang and Zhao, Yinghe and Liu, Lei and Li, Penghui and Nie, Anmin and Liu, Lixin and Yu, Jun and Feng, Xin and others},
  journal={Nature Materials},
  volume={22},
  number={9},
  pages={1078--1084},
  year={2023},
  publisher={Nature Publishing Group UK London}
}

@article{osada2012two,
  title={Two-Dimensional Dielectric Nanosheets: Novel Nanoelectronics from Nanocrystal Building Blocks},
  author={Osada, Minoru and Sasaki, Takayoshi},
  journal={Advanced Materials},
  volume={24},
  number={2},
  pages={210--228},
  year={2012},
  publisher={Wiley Online Library}
}

@article{lee2015highly,
  title={Highly Stable, Dual-Gated {MoS\textsubscript{2}} Transistors Encapsulated by Hexagonal Boron Nitride with Gate-Controllable Contact, Resistance, and Threshold Voltage},
  author={Lee, Gwan-Hyoung and Cui, Xu and Kim, Young Duck and Arefe, Ghidewon and Zhang, Xian and Lee, Chul-Ho and Ye, Fan and Watanabe, Kenji and Taniguchi, Takashi and Kim, Philip and others},
  journal={ACS Nano},
  volume={9},
  number={7},
  pages={7019--7026},
  year={2015},
  publisher={ACS Publications}
}

@article{yoon2025enabling,
  title={Enabling the Angstrom Era: {2D} Material-Based Multi-Bridge-Channel Complementary Field Effect Transistors},
  author={Yoon, Hoon Hahn and Park, Jin Young and Megra, Yonas Tsegaye and Baek, Ju Hwan and Song, Minuk and Akinwande, Deji and Ha, Daewon and Kang, Dong-Ho and Shin, Hyeon-Jin},
  journal={npj 2D Materials and Applications},
  volume={9},
  number={1},
  pages={68},
  year={2025},
  publisher={Nature Publishing Group UK London}
}

@article{shin20252d,
  title={{2D} Materials in Logic Technology: Power Efficiency and Scalability in {2DM-MBC CFET}},
  author={Shin, Seung Heon and Kang, Dong-Ho and Yoon, Hoon Hahn and Park, Jin Young and Song, Minuk and Son, Hyeonchang and Ha, Daewon and Shin, Hyeon-Jin},
  journal={Nano Letters},
  volume={25},
  number={18},
  pages={7224--7233},
  year={2025},
  publisher={ACS Publications}
}

@article{jung2025advances,
  title={Advances in Gate Dielectrics for {2D} Electronics},
  author={Jung, Moon-Chul and Kim, Jiwoo and Lee, Gwan-Hyoung},
  journal={physica status solidi (RRL)--Rapid Research Letters},
  volume={20},
  pages={2500133},
  year={2025},
  publisher={Wiley Online Library}
}

@article{pesin2012spintronics,
  title={Spintronics and Pseudospintronics in Graphene and Topological Insulators},
  author={Pesin, Dmytro and MacDonald, Allan H},
  journal={Nature Materials},
  volume={11},
  number={5},
  pages={409--416},
  year={2012},
  publisher={Nature Publishing Group UK London}
}

@article{chen2009experimental,
  title={Experimental Realization of a Three-Dimensional Topological Insulator, {Bi$_2$Te$_3$}},
  author={Chen, YL and Analytis, James G and Chu, J-H and Liu, ZK and Mo, S-K and Qi, Xiao-Liang and Zhang, HJ and Lu, DH and Dai, Xi and Fang, Zhong and others},
  journal={science},
  volume={325},
  number={5937},
  pages={178--181},
  year={2009},
  publisher={American Association for the Advancement of Science}
}

@article{moore2010birth,
  title={The Birth of Topological Insulators},
  author={Moore, Joel E},
  journal={Nature},
  volume={464},
  number={7286},
  pages={194--198},
  year={2010},
  publisher={Nature Publishing Group UK London}
}

@article{breunig2022opportunities,
  title={Opportunities in Topological Insulator Devices},
  author={Breunig, Oliver and Ando, Yoichi},
  journal={Nature Reviews Physics},
  volume={4},
  number={3},
  pages={184--193},
  year={2022},
  publisher={Nature Publishing Group UK London}
}

@article{cao2023future,
  title={The Future Transistors},
  author={Cao, Wei and Bu, Huiming and Vinet, Maud and Cao, Min and Takagi, Shinichi and Hwang, Sungwoo and Ghani, Tahir and Banerjee, Kaustav},
  journal={Nature},
  volume={620},
  number={7974},
  pages={501--515},
  year={2023},
  publisher={Nature Publishing Group UK London}
}

@article{jung2026advances,
  title={Advances and future challenges in monolithic 3d integrated logic, power, and optoelectronics technologies for tightly interconnected intelligent systems},
  author={Jung, Haksoon and Choi, Joonghoon and Baek, Seunghun and Shin, Bong Gyu and Song, Young Jae and Jung, Hanggyo and Kim, JoHyeon and Jeon, Jongwook and Kim, Gyumin and Park, Heechun and others},
  journal={ACS nano},
  volume={20},
  number={8},
  pages={6407--6445},
  year={2026},
  publisher={ACS Publications}
}

@article{castagne1971interface,
  title={Description of the {SiO$_2$}--{Si} interface properties by means of very low frequency {MOS} capacitance measurements},
  author={Castagne, R and Vapaille, A},
  journal={Surface Science},
  volume={28},
  number={1},
  pages={157--193},
  year={1971},
  publisher={Elsevier}
}

@article{yu2017photogenerated,
  title={Photogenerated charge harvesting and recombination in photocathodes of solvent-exfoliated WSe2},
  author={Yu, Xiaoyun and Sivula, Kevin},
  journal={Chemistry of Materials},
  volume={29},
  number={16},
  pages={6863--6875},
  year={2017},
  publisher={ACS Publications}
}

@article{kim2026optimizing,
  title={Optimizing cross-domain transfer for universal machine learning interatomic potentials},
  author={Kim, Jaesun and You, Jinmu and Park, Yutack and Lim, Yunsung and Kang, Yujin and Kim, Jisu and Jeon, Haekwan and Ju, Suyeon and Hong, Deokgi and Lee, Seung Yul and others},
  journal={Nature Communications},
  volume={17},
  number={1},
  pages={3432},
  year={2026},
  publisher={Nature Publishing Group UK London}
}

@inproceedings{li2004selected,
  title={Selected topics on {HfO$_2$} gate dielectrics for future {ULSI CMOS} devices},
  author={Li, MF and Yu, HY and Hou, YT and Kang, JF and Wang, XP and Shen, C and Ren, C and Yeo, YC and Zhu, CX and Chan, DSH and others},
  booktitle={Proceedings. 7th International Conference on Solid-State and Integrated Circuits Technology, 2004.},
  volume={1},
  pages={366--371},
  year={2004},
  organization={IEEE}
}

@article{engelherbert2010interface,
  title={Comparison of methods to quantify interface trap densities at dielectric/III-V semiconductor interfaces},
  author={Engel-Herbert, Roman and Hwang, Yoontae and Stemmer, Susanne},
  journal={Journal of applied physics},
  volume={108},
  number={12},
  year={2010},
  publisher={AIP Publishing}
}

@article{simmons1965richardson,
  title={Richardson-Schottky effect in solids},
  author={Simmons, JG},
  journal={Physical Review Letters},
  volume={15},
  number={25},
  pages={967},
  year={1965},
  publisher={APS}
}

@article{lenzlinger1969fowler,
  title={Fowler-Nordheim tunneling into thermally grown SiO2},
  author={Lenzlinger, M and Snow, EH},
  journal={Journal of Applied physics},
  volume={40},
  number={1},
  pages={278--283},
  year={1969},
  publisher={American Institute of Physics}
}

@article{chiu2014conduction,
  title={A review on conduction mechanisms in dielectric films},
  author={Chiu, Fu-Chien},
  journal={Advances in Materials Science and Engineering},
  volume={2014},
  number={1},
  pages={578168},
  year={2014},
  publisher={Wiley Online Library}
}

@article{yue2024topological,
  title={Topological Surface State Evolution in Bi2Se3 via Surface Etching},
  author={Yue, Ziqin and Huang, Jianwei and Wang, Ruohan and Li, Jia-Wan and Rong, Hongtao and Guo, Yucheng and Wu, Han and Zhang, Yichen and Kono, Junichiro and Zhou, Xingjiang and others},
  journal={Nano Letters},
  volume={24},
  number={40},
  pages={12413--12419},
  year={2024},
  publisher={ACS Publications}
}

@article{winter2013new,
  title={New method for determining flat-band voltage in high mobility semiconductors},
  author={Winter, Roy and Ahn, Jaesoo and McIntyre, Paul C and Eizenberg, Moshe},
  journal={Journal of Vacuum Science \& Technology B},
  volume={31},
  number={3},
  pages={030604},
  year={2013},
  publisher={AIP Publishing}
}

@article{hu2026ultrahigh,
  title={Ultrahigh Dielectric Permittivity in Ultrathin {2D $\beta$-Ga\textsubscript{2}O\textsubscript{3}} for Advanced Dielectric Applications},
  author={Hu, Xianyu and Liu, Zixiong and Wang, Xinglong and Guo, Qing and Feng, Xiyuan and Zhong, Yunlei},
  journal={Nano Letters},
  year={2026},
  volume={26},
  pages={1805--1812},
  publisher={ACS Publications}
}

@article{xu2022few,
  title={Few-layered {MnAl\textsubscript{2}S\textsubscript{4}} dielectrics for high-performance van der {Waals} stacked transistors},
  author={Xu, Fang and Wu, Ziyu and Liu, Guangjian and Chen, Feng and Guo, Junqing and Zhou, Hua and Huang, Jiawei and Zhang, Zhouyang and Fei, Linfeng and Liao, Xiaxia and others},
  journal={ACS applied materials \& interfaces},
  volume={14},
  number={22},
  pages={25920--25927},
  year={2022},
  publisher={ACS Publications}
}

@article{chang2021ald,
  title={{ALD-{ZrO\textsubscript{2}}} gate dielectric with suppressed interfacial oxidation for high performance {MoS\textsubscript{2}} top gate {MOSFETs}},
  author={Chang, Wen Hsin and Okada, Naoya and Horikawa, Masayo and Endo, Takahiko and Miyata, Yasumitsu and Irisawa, Toshifumi},
  journal={Japanese Journal of Applied Physics},
  volume={60},
  number={SB},
  pages={SBBH03},
  year={2021},
  publisher={IOP Publishing}
}

@article{fu2025low,
  title={Low-temperature controlled growth of {2D LaOCl} with enhanced dielectric properties for advanced electronics},
  author={Fu, Zhipeng and Jian, Chuanyong and Yao, Yu and Li, Yixiang and Yuan, Jiashuai and Cai, Qian and Liu, Wei},
  journal={Advanced Functional Materials},
  volume={35},
  number={34},
  pages={2501136},
  year={2025},
  publisher={Wiley Online Library}
}

@article{wen2016effects,
  title={Effects of annealing on electrical performance of multilayer {MoS\textsubscript{2}} transistors with atomic layer deposited {HfO\textsubscript{2}} gate dielectric},
  author={Wen, Ming and Xu, Jingping and Liu, Lu and Lai, Pui-To and Tang, Wing-Man},
  journal={Applied Physics Express},
  volume={9},
  number={9},
  pages={095202},
  year={2016},
  publisher={The Japan Society of Applied Physics}
}

@article{zou2019improved,
  title={Improved performance of top-gated multilayer {MoS\textsubscript{2}} transistors with channel fully encapsulated by {Al\textsubscript{2}O\textsubscript{3}} dielectric},
  author={Zou, Jiyue and Wang, Lisheng and Chen, Fengxiang},
  journal={AIP Advances},
  volume={9},
  number={9},
  pages={095061},
  year={2019},
  publisher={AIP Publishing}
}

@article{ziegler2010srim,
  title={SRIM--The stopping and range of ions in matter (2010)},
  author={Ziegler, James F and Ziegler, Matthias D and Biersack, Jochen P},
  journal={Nuclear Instruments and Methods in Physics Research Section B: Beam Interactions with Materials and Atoms},
  volume={268},
  number={11-12},
  pages={1818--1823},
  year={2010},
  publisher={Elsevier}
}

@article{konobeyev2017evaluation,
  title={Evaluation of effective threshold displacement energies and other data required for the calculation of advanced atomic displacement cross-sections},
  author={Konobeyev, A Yu and Fischer, U and Korovin, Yu A and Simakov, SP},
  journal={Nuclear Energy and Technology},
  volume={3},
  number={3},
  pages={169--175},
  year={2017},
  publisher={Elsevier}
}

@article{pourfath2026device,
  title={Device-scaling constraints imposed by the van der Waals gap formed in two-dimensional materials},
  author={Pourfath, Mahdi and Grasser, Tibor},
  journal={Science},
  volume={392},
  pages={eaeb2271},
  year={2026},
  publisher={American Association for the Advancement of Science}
}

@article{kong2011rapid,
  title={{Rapid Surface Oxidation as a Source of Surface Degradation Factor for Bi\textsubscript{2}Se\textsubscript{3}}},
  author={Kong, Desheng and Cha, Judy J and Lai, Keji and Peng, Hailin and Analytis, James G and Meister, Stefan and Chen, Yulin and Zhang, Hai-Jun and Fisher, Ian R and Shen, Zhi-Xun and others},
  journal={ACS nano},
  volume={5},
  number={6},
  pages={4698--4703},
  year={2011},
  publisher={ACS Publications}
}

@article{lieberman1994principles,
  title={Principles of plasma discharges and materials processing},
  author={Lieberman, Michael A and Lichtenberg, Allan J},
  journal={MRS Bulletin},
  volume={30},
  number={12},
  pages={899--901},
  year={1994}
}

@article{kawamura1999ion,
  title={Ion energy distributions in rf sheaths; review, analysis and simulation},
  author = {Kawamura, E and Vahedi, V and Lieberman, MA and Birdsall, CK},
  journal={Plasma Sources Science and Technology},
  volume={8},
  number={3},
  pages={R45--R64},
  year={1999}
}

@article{gahan2012ion,
  title={Ion energy distribution measurements in rf and pulsed dc plasma discharges},
  author={Gahan, D and Daniels, S and Hayden, C and Scullin, P and O'sullivan, D and Pei, YT and Hopkins, MB},
  journal={Plasma Sources Science and Technology},
  volume={21},
  number={2},
  pages={024004},
  year={2012},
  publisher={IoP Publishing}
}

@article{franklin2002electronegative,
  title={{Electronegative plasmas—Why are they so different?}},
  author={Franklin, RN},
  journal={Plasma Sources Science and Technology},
  volume={11},
  number={3A},
  pages={A31--A37},
  year={2002}
}

@article{kono2002negative,
  title={Negative ions in processing plasmas and their effect on the plasma structure},
  author={Kono, Akihiro},
  journal={Applied Surface Science},
  volume={192},
  number={1-4},
  pages={115--134},
  year={2002},
  publisher={Elsevier}
}

@article{kresse1996efficient,
  title={Efficient iterative schemes for ab initio total-energy calculations using a plane-wave basis set},
  author={Kresse, Georg and Furthm{\"u}ller, J{\"u}rgen},
  journal={Physical review B},
  volume={54},
  number={16},
  pages={11169},
  year={1996},
  publisher={APS}
}

@article{perdew1996generalized,
  title={Generalized gradient approximation made simple},
  author={Perdew, John P and Burke, Kieron and Ernzerhof, Matthias},
  journal={Physical review letters},
  volume={77},
  number={18},
  pages={3865},
  year={1996},
  publisher={APS}
}

@article{blochl1994projector,
  title={Projector augmented-wave method},
  author={Bl{\"o}chl, Peter E},
  journal={Physical review B},
  volume={50},
  number={24},
  pages={17953},
  year={1994},
  publisher={APS}
}

@article{gonze1997dynamical,
  title={Dynamical matrices, Born effective charges, dielectric permittivity tensors, and interatomic force constants from density-functional perturbation theory},
  author={Gonze, Xavier and Lee, Changyol},
  journal={Physical Review B},
  volume={55},
  number={16},
  pages={10355},
  year={1997},
  publisher={APS}
}

@article{grimme2011effect,
  title={Effect of the damping function in dispersion corrected density functional theory},
  author={Grimme, Stefan and Ehrlich, Stephan and Goerigk, Lars},
  journal={Journal of computational chemistry},
  volume={32},
  number={7},
  pages={1456--1465},
  year={2011},
  publisher={Wiley Online Library}
}

@article{shen2025mos2,
  title={{MoS\textsubscript{2} transistors with 4 nm hBN gate dielectric and 0.46 V threshold voltage}},
  author={Shen, Yaqing and Pazos, Sebastian and Zheng, Wenwen and Yuan, Yue and Ping, Yue and Alharbi, Osamah and Liu, Hang and Lu, Xu and Lanza, Mario},
  journal={ACS nano},
  volume={19},
  number={17},
  pages={16903--16912},
  year={2025},
  publisher={ACS Publications}
}

\end{document}